\documentclass{mytlp}

\usepackage{amsmath}
\usepackage{amssymb}
\usepackage{harvard}

\allowdisplaybreaks[1]

\newcommand{\summary}[1]{\textrm{\textbf{\textup{#1}}}}

\renewcommand{\emptyset}{\mathord{\varnothing}}

\newcommand*{\wpf}{\mathop{\wp_\mathrm{f}}\nolimits}

\newcommand*{\sseq}{\subseteq}

\newcommand*{\Sseq}{\supseteq}

\newcommand*{\union}{\mathbin{\cup}}
\newcommand*{\inters}{\mathbin{\cap}}
\newcommand*{\setdiff}{\mathbin{\setminus}}
\newcommand*{\bigunion}{\bigcup}

\newcommand*{\Vars}{\mathord{\mathit{Vars}}}
\newcommand*{\vars}{\mathop{\mathit{vars}}\nolimits}

\providecommand*{\Nset}{\mathbb{N}}            

\newcommand{\lub}{\mathop{\mathrm{lub}}\nolimits}

\newcommand*{\fund}[3]{\mathord{#1}\colon#2\rightarrow#3}
\newcommand*{\reld}[3]{\mathord{#1}\subseteq#2\times#3}

\newcommand*{\dom}{\mathop{\mathrm{dom}}\nolimits}
\newcommand*{\param}{\mathop{\mathrm{param}}\nolimits}

\newcommand{\defrel}[1]{\mathrel{\buildrel \mathrm{def} \over {#1}}}
\newcommand{\defeq}{\defrel{=}}

\newcommand{\st}{\mathrel{.}}
\newcommand{\itc}{\mathrel{:}}

\newcommand*{\piff}{\mathrel{\leftrightarrow}}
\newcommand*{\pimplies}{\mathrel{\rightarrow}}
\newcommand*{\entails}{\mathrel{\vdash}}

\newcommand*{\Sharing}{\textup{\textsf{Sharing}}}
\newcommand*{\SG}{\mathit{SG}}
\newcommand*{\SH}{\mathit{SH}}
\newcommand*{\sh}{\mathit{sh}}
\newcommand*{\SSl}{\mathit{SS}}
\newcommand*{\leqSSl}{\mathrel{\preceq_\mathit{\scriptscriptstyle SS}}}
\newcommand*{\Def}{\mathit{Def}}

\newcommand{\china}{\textmd{\textsc{China}}}

\newcommand*{\cS}{\ensuremath{\mathcal{S}}}
\newcommand*{\cT}{\ensuremath{\mathcal{T}}}

\newcommand*{\Terms}{\cT_{\Vars}}
\newcommand*{\GTerms}{\cT_{\emptyset}}

\newcommand*{\PS}{\mathit{PS}}

\newcommand*{\Bind}{\mathit{Bind}}

\newcommand*{\RSubst}{\mathit{RSubst}}
\newcommand*{\VSubst}{\mathit{VSubst}}
\newcommand*{\ISubst}{\mathit{ISubst}}

\newcommand*{\rel}{\mathop{\mathrm{rel}}\nolimits}
\newcommand*{\irel}{\mathop{\overline{\mathrm{rel}}}\nolimits}
\newcommand*{\bin}{\mathop{\mathrm{bin}}\nolimits}

\newcommand*{\amgu}{\mathop{\mathrm{amgu}}\nolimits}
\newcommand*{\Amgu}{\mathop{\mathrm{Amgu}}\nolimits}
\newcommand*{\amge}{\mathop{\mathrm{amge}}\nolimits}
\newcommand*{\aunify}{\mathop{\mathrm{aunify}}\nolimits}

\newcommand*{\var}{\mathop{\mathrm{var}}\nolimits}
\newcommand*{\lhs}{\mathop{\mathrm{lhs}}\nolimits}
\newcommand*{\rhs}{\mathop{\mathrm{rhs}}\nolimits}
\newcommand*{\Eqs}{\mathop{\mathrm{Eqs}}\nolimits}

\newcommand{\Sstep}{\mathrel{\buildrel{\cS}\over\longmapsto}}
\newcommand{\Sstepstar}{\mathrel{\buildrel{\mkern-10mu\cS}\over{\longmapsto^{\smash{\mkern-2.2mu\ast}}}}}
 
\newcommand{\sg}{\mathop{\mathrm{sg}}\nolimits}
\newcommand{\occ}{\mathop{\mathrm{occ}}\nolimits}

\newcommand*{\compose}{\mathbin{\circ}}

\newcommand{\just}[1]{\text{[#1]}}

\newtheorem{thm}{Theorem}
\newtheorem{cor}{Corollary}

\newtheorem{lem}{Lemma}
\newtheorem{exmp}{Example}
\newtheorem{defn}{Definition}

\begin{document}
\title[Soundness, Idempotence and Commutativity of Set-Sharing]
      {Soundness, Idempotence and Commutativity \\ of Set-Sharing}
\author[P. M. Hill, R. Bagnara and E. Zaffanella]
       {PATRICIA M. HILL\thanks{This work was partly supported
                                 by EPSRC under grant GR/M05645.} \\
       School of Computer Studies,
       University of Leeds,
       Leeds, U.K. \authorbreak
       \email{hill@scs.leeds.ac.uk}
       \and ROBERTO BAGNARA, ENEA ZAFFANELLA\thanks{The work of the second
                          and third authors
                          has been partly supported by MURST project
                          ``Certificazione automatica di programmi
                            mediante interpretazione astratta.''} \\
       Department of Mathematics,
       University of Parma,
       Italy \authorbreak
       \email{\{bagnara,zaffanella\}@cs.unipr.it}
}
\maketitle

\begin{abstract}
It is important that practical data-flow analyzers are backed by
reliably proven theoretical results.
Abstract interpretation provides a sound mathematical framework
and necessary generic properties for an
abstract domain to be well-defined and sound with respect to the
concrete semantics.
In logic programming, the abstract domain $\Sharing$ is a standard choice
for sharing analysis for both practical work and further theoretical study.
 In spite of this, we found that there were no satisfactory proofs for the
key properties of commutativity and idempotence that are essential for
$\Sharing$ to be well-defined and that published statements
of the soundness of $\Sharing$ assume the occurs-check.
This paper provides a generalization of the abstraction function for
$\Sharing$
that can be applied to any language, with or without the occurs-check.
Results for soundness, idempotence and commutativity for
abstract unification using this abstraction function are proven.
\end{abstract}


\section{Introduction}

In abstract interpretation, the concrete semantics of a program is
approximated by an abstract semantics;
that is, the concrete domain is replaced by an abstract domain
and each elementary operation on the concrete domain is replaced by
a corresponding abstract operation on the abstract domain.
Assuming the global abstract procedure mimics
the concrete execution procedure,
each basic operation on the elements of the abstract domain
must produce a safe approximation of the corresponding operation
on corresponding elements of the concrete domain.
For logic programming, the key elementary operation is
\emph{unification} that computes a solution to a set of equations.
This solution can be represented by means of a mapping
(called a \emph{substitution}) from variables to first-order
terms in the language.
For global soundness of the abstract semantics,
there needs to be, therefore,
a corresponding abstract operation, \emph{aunify},
that is sound with respect to unification.

For parallelization and several other program
optimizations, it is important to know before execution
which variables may be bound to terms that share a common variable.
Jacobs and Langen developed the abstract domain
$\Sharing$ \cite{JacobsL89,JacobsL92}
for representing
and propagating 
the sharing behavior of variables and
this is now a standard choice for sharing analysis.
Subsequent research then concentrated mainly on extending the domain to
incorporate additional properties such as
linearity, freeness and depth-$k$ abstractions
\cite{Langen90th,BruynoogheC93,CodishDFB96,King94,KingS94,MuthukumarH92}
or in reducing its complexity~\cite{BagnaraHZ97b,BagnaraHZ01TCS}.
Key properties such as commutativity and soundness of this domain and its
associated abstract operations such as abstract unification were normally
assumed to hold.
One reason for this was that~\cite{JacobsL92} includes
a proof of the soundness and refers to the Ph.D.~thesis of Langen
\cite{Langen90th} for the proofs of commutativity
and idempotence.\footnote{Even though the thesis of Langen has been
published as a technical report of the University of Southern California,
an extensive survey of the literature on $\Sharing$ indicates that the
thesis has not been widely circulated even among researchers in the field.
For instance, Langen is rarely credited as being the first person to
integrate $\Sharing$ with linearity information, despite the fact that
this is described in the thesis.}
We discuss below why these results are inadequate.

\subsection{Soundness of $\aunify$}

An important step in standard unification algorithms
based on that of Robinson~\cite{Robinson65}
(such as the Martelli-Montanari algorithm~\cite{MartelliM82})
is the \emph{occurs-check},
which avoids the generation of infinite (or cyclic) data structures.
With such algorithms, the resulting solution is both unique and idempotent.
However, in computational terms, the occurs-check is expensive
and the vast majority of Prolog implementations omit this test,
although some Prolog implementations do offer unification with the occurs-check
as a separate built-in predicate
(in ISO Prolog~\cite{ISO-Prolog-part-1}
the predicate is \texttt{unify\_with\_occurs\_check/2}).
In addition, if the unification algorithm is based on the
Martelli-Montanari algorithm but without the occurs-check step,
then the resulting solution may be non-idempotent.
Consider the following example.

Suppose we are given as input the equation
$p\bigl(z,f(x,y)\bigr) = p\bigl(f(z,y),z\bigr)$
with an initial substitution that is empty.
We apply the steps in the Martelli-Montanari procedure but
without the occurs-check:
\begin{align*}
&&   \text{equations}          && \text{substitution}\\
\\
1 && p(z,f(x,y)) = p(f(z,y),z) && \emptyset\\
2 && z = f(z,y), f(x,y) = z    && \emptyset\\
3 && f(x,y) = f(z,y)           && \bigl\{z \mapsto f(z,y)\bigr\}\\
4 && x=z, y=y                  && \bigl\{z \mapsto f(z,y)\bigr\}\\
5 && y=y                       && \bigl\{z \mapsto f(z,y), x \mapsto z\bigr\}\\
6 && \emptyset                 && \bigl\{z \mapsto f(z,y), x \mapsto z\bigr\}
\end{align*}
Then
\(
  \sigma = \bigl\{ z \mapsto f(z,y), x \mapsto z \bigr\}
\)
is the computed substitution; it is not idempotent
since, for example, $x\sigma = z$ and $x\sigma\sigma = f(z,y)$.

Non-standard equality theories and unification procedures are
also available and used in many logic programming systems.
In particular, there are theoretically coherent languages,
such as  Prolog~III~\cite{Colmerauer82},
that employ an equality theory and unification algorithm
based on a theory of \emph{rational trees}
(possibly infinite trees with a finite number of subtrees).
As remarked in \cite{Colmerauer82}, complete (i.e., always terminating)
unification with the omission of the occurs-check solves equations
over rational trees. Complete unification is made available by several
Prolog implementations.
The substitutions computed by such systems are
in \emph{rational solved form}
and therefore not necessarily idempotent.
As an example, the substitution $\{x \mapsto f(x)\}$,
which is clearly non-idempotent,
is in rational solved form
and could itself be computed by the above algorithms.

It is therefore important that theoretical work in data-flow analysis
makes no assumption that the computed solutions are idempotent.
In spite of this, most theoretical work on data-flow analysis of
logic programming \emph{and} of Prolog assume the occurs-check
is performed, thus allowing idempotent substitutions only.
In particular, \cite{JacobsL92}, \cite{Langen90th}, and, more recently,
\cite{CortesiF99} make this assumption in their proofs of soundness.
As a consequence, their results do not apply to the analysis of all Prolog
programs. 
A recent exception to this is~\cite{King00}
where a soundness result is proved for a domain
representing just the pair-sharing and linearity information.
In this work it is assumed that a separate groundness analysis
is performed and its results are used to recover from
the precision losses incurred by the proposed domain.
However, the problem of specifying a sound and precise groundness
analysis when dealing with possibly non-idempotent substitutions
is completely disregarded,
so that the overall solution is incomplete.
Moreover, the proposed abstraction function is based on a limit
operation that, in the general case, is not finitely computable.

We have therefore addressed the problem of defining a sound and precise
approximation of the sharing information contained in a substitution
in rational solved form.

In particular, we observed that the $\Sharing$ domain
is concerned with the set of variables occurring in a term,
rather than with the term structure.
We have therefore generalized the notion of idempotence
to \emph{variable-idempotence}.
That is, if $\sigma$ is a variable-idempotent substitution
and $t$ is any term,
then any variable which is not in the domain of $\sigma$
and occurs in $t\sigma\sigma$ also occurs in $t\sigma$.
Clearly, as illustrated by the above example,
substitutions generated by unification algorithms
without the occurs-check may not even be variable-idempotent.
To resolve this,
we have devised an algorithm that transforms any substitution
in rational solved form to an equivalent
(with respect to any equality theory)
variable-idempotent substitution.
For instance, in the example, it would transform $\sigma$ to
$\bigl\{z \mapsto f(z, y), x \mapsto f(z,y) \bigr\}$.

By suitably exploiting the properties enjoyed by variable-idempotent
substitutions, we show that, for the domain $\Sharing$,
the abstract unification algorithm $\aunify$ is sound
with respect to the actually implemented unification procedures
for all logic programming languages.
Moreover, we define a new abstraction function mapping any set of
substitutions in rational solved form into the corresponding
abstract descriptions so that there is no need for the analyser to
compute the equivalent set of variable-idempotent substitutions.
We note that this new abstraction function is carefully chosen 
so as to avoid any precision loss due to the possible 
non-idempotence of the substitution.

Note that both the notion of variable-idempotent substitution and the
proven results relating it to arbitrary substitutions in rational
solved form do not depend on the particular abstract domain
considered.  Indeed, we believe that this concept, perhaps with
minor adjustments, can be usefully applied to other abstract domains
when extending the soundness proofs devised for idempotent
substitutions to the more general case of substitutions in rational
solved form.

\subsection{Commutativity and Idempotence of $\aunify$}

A substitution is defined as a \emph{set} of bindings or
equations between variables and other terms.
Thus, for the concrete domain, the order and multiplicity of elements
are irrelevant in both the computation and semantics of unification.
It is therefore useful that the abstraction of the unification procedure
should be unaffected by the order and multiplicity in which it abstracts
the bindings that are present in the substitution.
Furthermore, from a practical perspective, it is also useful if the global
abstract procedure can proceed in a different order
with respect to the concrete one
without affecting the accuracy of the analysis results.
On the other hand, as sharing is normally combined with linearity
and freeness domains that are not idempotent or
commutative~\cite{Langen90th,BruynoogheC93,King94},
it may be asked why these properties are still important for sharing analysis.
In answer to this, we observe that the order and multiplicity in which
the bindings in a substitution are analyzed affects the accuracy of
the linearity and freeness information.
It is therefore a real advantage to be able to ignore these aspects
as far as the sharing domain is concerned.
Specifically, the order in which the bindings are analyzed can be chosen
so as to improve the accuracy of linearity and freeness.
We thus conclude that it is extremely desirable that $\aunify$ is also
commutative and idempotent.

We found that there was no satisfactory proof of commutativity.
In addition, for idempotence the only previous result was
given in~\cite[Theorem 32]{Langen90th} of the thesis of Langen.
However, his definition of abstract unification
includes the renaming and
projection operations and, in this case, only a weak form of
idempotence holds. In fact, for the basic $\aunify$
operation as defined here and without projection and renaming,
idempotence has never before been proven.
We therefore provide here the first published proofs of these properties.

In summary, this paper,
which is an extended and improved version of \cite{HillBZ98b},
provides a generalization of the abstraction function for $\Sharing$
that can be applied to any logic programming language
dealing with syntactic term structures.
The results for soundness, idempotence and commutativity for
abstract unification using this abstraction function are proved.

The paper is organised as follows.
In the next section, the notation and definitions needed
for equality and substitutions in the concrete domain are given.
In Section~\ref{sec: set sharing}, we recall the
definition of the domain $\Sharing$ and of the classical
abstraction function defined for idempotent substitutions.
We also show why this abstraction function cannot be applied,
as is, to non-idempotent substitutions.
In Section~\ref{sec:Variable-Idempotence}, we introduce
\emph{variable-idempotence} and provide a transformation that may be used to 
map any substitution in rational solved form to an 
equivalent, variable-idempontent one.
In Section~\ref{sec:abstracting-in-rsubst}, we define
a new abstraction function relating the $\Sharing$ domain
to the domain of arbitrary substitutions in rational solved form.
In Section~\ref{sec: abstract unify},
we recall the definition of the abstract unification
for $\Sharing$ and state our main results.
Section~\ref{sec: discuss} concludes.
For the convenience of the reader, throughout the paper
all the proofs (apart from the simpler ones) of the stated results
are appended to the end of the corresponding section.

\section{Equations and Substitutions}
\label{sec:preliminaries}

In this section we introduce the notation and some terminology
concerning equality and substitutions that will be used
in the rest of the paper.

\subsection{Notation}

For a set $S$,
$\wp(S)$ is the powerset of $S$,
whereas $\wpf(S)$ is the set of all the \emph{finite} subsets of $S$.
The symbol $\Vars$ denotes a denumerable set of variables,
whereas $\Terms$ denotes the set of first-order terms over $\Vars$
for some given set of function symbols.
It is assumed that there are at least two distinct function symbols,
one of which is a constant (i.e., of zero arity), in the given set.
The set of variables occurring in a syntactic object $o$ is denoted
by $\vars(o)$.
To simplify the expressions in the paper,
any variable in a formula that is not in the scope of a quantifier
is assumed to be universally quantified.
To prove the results in the paper,
it is useful to assume a total ordering,
denoted with `$\leq$', on $\Vars$.

\subsection{Substitutions}

A \emph{substitution} is a total function $\fund{\sigma}{\Vars}{\Terms}$
that is the identity almost everywhere;
in other words, the \emph{domain} of~$\sigma$,
\[
  \dom(\sigma) \defeq \bigl\{\, x \in \Vars \bigm| \sigma(x) \neq x \,\bigr\},
\]
is finite.
Given a substitution $\fund{\sigma}{\Vars}{\Terms}$
we overload the symbol `$\sigma$' so as to denote also
the function $\fund{\sigma}{\Terms}{\Terms}$ defined
as follows, for each term $t \in \Terms$:
\[
  \sigma(t)
    \defeq
      \begin{cases}
        t,          &\text{if $t$ is a constant symbol;} \\
        \sigma(t),  &\text{if $t \in \Vars$;} \\
        f\bigl(\sigma(t_1), \ldots, \sigma(t_n)\bigr),
                   &\text{if $t = f(t_1, \ldots, t_n)$.}
    \end{cases}
\]
If $t \in \Terms$, we write $t \sigma$ to denote $\sigma(t)$
and $t[x/s]$ to denote $t\{x \mapsto s\}$.

If $x\in\Vars$ and $s\in\Terms\setdiff\{x\}$,
then $x\mapsto s$ is called a \emph{binding}.
The set of all bindings is denoted by $\Bind$.
Substitutions are syntactically denoted by the set of their bindings,
thus a substitution $\sigma$ is identified with the (finite) set
\[
  \bigl\{\, x \mapsto \sigma(x) \bigm| x \in \dom(\sigma) \,\bigr\}.
\]
Thus, $\vars(\sigma)$ is the set of variables occurring in the
bindings of $\sigma$ and we also define the set of \emph{parameter variables}
of a substitution $\sigma$ as
\[
  \param(\sigma) \defeq \vars(\sigma) \setminus \dom(\sigma).
\]

A substitution is said to be \emph{circular} if, for $n > 1$, it has the
form
\[
  \{x_1\mapsto x_2, \ldots, x_{n-1}\mapsto x_n, x_n\mapsto x_1\},
\]
where $x_1$, \ldots, $x_n$ are distinct variables.
A substitution is in \emph{rational solved form} if it has no circular subset.
The set of all substitutions in rational solved form is denoted by
$\RSubst$.
A substitution $\sigma$ is \emph{idempotent} if, for all $t\in\Terms$,
we have $t\sigma\sigma=t\sigma$. The set of all idempotent substitutions
is denoted by $\ISubst$ and $\ISubst \subset \RSubst$.

\begin{exmp}
\label{ex: rat solved form}
The following hold:
\begin{align*}
  \bigl\{ x \mapsto y, y \mapsto a \bigr\} & \in \RSubst \setminus \ISubst,\\
  \bigl\{ x \mapsto a, y \mapsto a \bigr\} & \in \ISubst,\\
  \bigl\{ x \mapsto y, y \mapsto g(y) \bigr\} & \in \RSubst \setminus \ISubst,\\
  \bigl\{ x \mapsto y, y \mapsto g(x) \bigr\} & \in \RSubst \setminus \ISubst,\\
  \bigl\{ x \mapsto y, y \mapsto x \bigr\} & \notin \RSubst, \\
  \bigl\{ x \mapsto y, y \mapsto x, z\mapsto a \bigr\} & \notin \RSubst.
\end{align*}
\end{exmp}

We have assumed that there is a total ordering `$\leq$' for $\Vars$.
We say that $\sigma\in \RSubst$ is \emph{ordered}
(with respect to this ordering) if,
for each binding $(v\mapsto w) \in \sigma$ such that
$w \in \param(\sigma)$, we have $w < v$.

The composition of substitutions is defined in the usual way.
Thus $\tau \compose \sigma$ is the substitution such that,
for all terms $t \in \Terms$,
\[
  (\tau\compose \sigma)(t) = \tau\bigl(\sigma(t)\bigr)
\]
and has the formulation
\begin{equation}
\label{eq:alt-compose}
  \tau\compose \sigma
    =
      \bigl\{\,
        x \mapsto x\sigma\tau 
      \bigm|
        x \in \dom(\sigma),
        x \neq x\sigma\tau
      \,\bigr\}
        \union 
          \bigl\{\,
            x \mapsto x\tau
          \bigm|
            x \in \dom(\tau) \setdiff \dom(\sigma)   
          \,\bigr\}.
\end{equation}
As usual, $\sigma^0$ denotes
the identity function (i.e., the empty substitution)
and, when $i > 0$, 
$\sigma^i$ denotes the substitution $(\sigma\circ\sigma^{i-1})$.

\subsection{Equations}

An \emph{equation} is of the form $s = t$ where $s,t\in\Terms$.
$\Eqs$ denotes the set of all equations.
A substitution $\sigma$
may be regarded as a finite set of equations,
that is, as the set
$\bigl\{\, x = t \bigm| (x \mapsto t) \in \sigma \,\bigr\}$.
We say that a set of equations $e$ is in \emph{rational solved form}
if $\bigl\{\, s \mapsto t \bigm| (s=t) \in e \,\bigr\} \in \RSubst$.
In the rest of the paper, we will often write
a substitution $\sigma \in \RSubst$
to denote a set of equations in rational solved form
(and vice versa).

We assume that any equality theory $T$ over $\Terms$
includes the \emph{congruence axioms}
denoted by the following schemata:
\begin{align}
\label{eq-ax:id}
s=s &,  \\
\label{eq-ax:sym}
s=t &\piff t=s, \\
\label{eq-ax:trans}
r=s \land s=t &\pimplies r=t, \\
\label{eq-ax:congr}
s_1=t_1 \land \cdots \land s_n=t_n
  &\pimplies
    f(s_1, \ldots, s_n) = f(t_1, \ldots, t_n).
\end{align}

In logic programming and most implementations of Prolog
it is usual to assume an equality theory based on syntactic identity.
This consists of the congruence axioms
together with the \emph{identity axioms} denoted by
the following schemata,
where $f$ and $g$ are distinct function symbols or $n \neq m$:
\begin{gather}
\label{eq-ax:injective-functions}
f(s_1, \ldots, s_n) = f(t_1, \ldots, t_n)
  \pimplies
    s_1=t_1 \land \cdots \land s_n=t_n, \\
\label{eq-ax:diff-funct}
\neg \bigl(f(s_1,\ldots,s_n) = g(t_1,\ldots,t_m)\bigr).
\end{gather}
The axioms characterized by
schemata~(\ref{eq-ax:injective-functions}) and~(\ref{eq-ax:diff-funct})
ensure the equality theory depends only on the syntax.
The equality theory for a non-syntactic domain replaces these axioms
by ones that depend instead on the semantics of the domain and,
in particular, on the interpretation given to functor symbols.

The equality theory of Clark~\cite{Clark78}
on which pure logic programming is based,
usually called the \emph{Herbrand} equality theory,
is given by the congruence axioms, the identity axioms,
and the axiom schema
\begin{equation}
\label{eq-ax:occ-check}
 \forall z\in \Vars \itc
    \forall t \in (\Terms\setdiff\Vars) \itc
      z \in \vars(t) \pimplies \neg (z = t).
\end{equation}
Axioms characterized by the schema~(\ref{eq-ax:occ-check})
are called the \emph{occurs-check axioms} and are an essential
part of the standard unification procedure in SLD-resolution.

An alternative approach used in some implementations
of Prolog,
does not require the occurs-check axioms.
This approach is based on the theory of
rational trees~\cite{Colmerauer82,Colmerauer84}.
It assumes the congruence axioms and the identity axioms together with a
\emph{uniqueness axiom} for each substitution in rational solved form.
Informally speaking these state that,
after assigning a ground rational tree to each parameter variable,
the substitution uniquely defines a ground rational tree
for each of its domain variables.
Note that being in rational solved form is a very weak property.
Indeed, unification algorithms returning a set of equations
in rational solved form are allowed to be much more ``lazy''
than one would usually expect
(e.g., see the first substitution in Example~\ref{ex: rat solved form}).
We refer the interested reader to \cite{JaffarLM87,Keisu94th,Maher88}
for details on the subject.

In the sequel we will use the expression ``equality theory''
to denote any consistent, decidable theory $T$ satisfying the congruence
axioms.
We will also use the expression ``syntactic equality theory'' to denote
any equality theory $T$ also satisfying
the identity axioms.\footnote{Note that, as a consequence of
axiom~(\ref{eq-ax:diff-funct})
and the assumption that there are at least two
distinct function symbols in the language, one of which is a constant,
there exist two terms $a_1, a_2 \in \GTerms$ such that,
for any syntactic equality theory $T$, we have $T \entails a_1 \neq a_2$.}
When the equality theory $T$ is clear from the context, 
it is convenient to adopt the notations $\sigma \implies \tau$ and 
$\sigma \iff \tau$, where $\sigma,\tau$ are sets of equations, to denote 
$T\entails \forall(\sigma \pimplies \tau)$ and 
$T\entails \forall(\sigma \piff \tau)$, respectively.

Given an equality theory $T$, and a set of equations in rational solved form
$\sigma$, we say that $\sigma$ is \emph{satisfiable} in $T$ if
$T \entails
\forall \Vars\setdiff\dom(\sigma)\itc \exists \dom(\sigma)
\st \sigma$.
If $T$ is a syntactic equality
theory that also includes the occurs-check axioms,
and $\sigma$ is satisfiable in $T$,
then we say that $\sigma$ is \emph{Herbrand}.

Given a satisfiable set of equations $e\in\wpf(\Eqs)$ 
in an equality theory $T$,
then a substitution $\sigma \in \RSubst$ is called a 
\emph{solution for $e$ in $T$}
if $\sigma$ is satisfiable in $T$ and
$T \entails \forall(\sigma \pimplies e)$.
If $\vars(\sigma) \sseq \vars(e)$, then
$\sigma$ is said to be a \emph{relevant} solution for $e$.
In addition, $\sigma$ is a
\emph{most general solution for $e$ in $T$} if
$T \entails \forall(\sigma \piff e)$.
In this paper, a most general solution is always a relevant solution of $e$.

Observe that, given an equality theory $T$, a set of equations
in rational solved form may not be satisfiable in $T$.
For example, $\exists x \itc \bigl\{x = f(x)\bigr\}$ is false
in the Clark equality theory.

\begin{lem}
\label{lem: add-binding}
Suppose $T$ is an equality theory, $\sigma \in \RSubst$ is satisfiable in $T$,
$x \in \Vars \setdiff \dom(\sigma)$, and $a \in \GTerms$.
Then, $\sigma' \defeq \sigma \union \{x \mapsto a\} \in \RSubst$
and $\sigma'$ is satisfiable in $T$.
\end{lem}
\begin{proof}
As $x \notin \dom(\sigma)$ and $\sigma \in \RSubst$ and $a \in \GTerms$,
it follows that $\sigma' = \sigma \union \{x \mapsto a\} \in \RSubst$.

Since $\sigma$ is satisfiable in $T$,
\begin{align*}
  T
    &\entails
        \forall \Vars \setdiff \dom(\sigma)
          \itc \exists \dom(\sigma) \st \sigma. \\
\intertext{%
Moreover, by the congruence axiom~(\ref{eq-ax:id}),
}
  T
    &\entails
      \forall \Vars \setdiff \{x\} \itc \exists x \st \{x = a\}. \\
\intertext{%
Hence,
}
  T
    &\entails
      \forall \Vars \setdiff \bigl(\dom(\sigma) \union \{x\}\bigr)
        \itc \exists \bigl(\dom(\sigma) \union \{x\} \bigr)
               \st \sigma \union \{x = a\}.
\end{align*}
Thus $\sigma' = \sigma \union \{x \mapsto a\}$ is satisfiable in $T$.
\end{proof}

Syntactically we have shown that any substitution in $\RSubst$ may be regarded
as a set of equations in rational solved form and vice versa.
The next lemma shows the semantic relationship between them.

\begin{lem}
\label{lem: applic}
If $T$ is an equality theory and $\sigma \in \RSubst$, then,
for each $t \in \Terms$,
\[
  T
    \entails
      \forall \bigl(\sigma \pimplies (t=t\sigma)\bigr).
\]
\end{lem}
\begin{proof}
We assume the congruence axioms hold and prove that,
for any $t\in \Terms $, we have $\sigma \implies \{t=t\sigma\}$.
The proof is by induction on the depth of $t$.

Suppose, first that the depth of $t$ is one.
If $t$ is a variable not in $\dom(\sigma)$ or a constant, then
$t\sigma = t$ and the result follows from axiom (\ref{eq-ax:id}).
If $t \in \dom(\sigma)$, then,
for some $r \in \Terms$, $(t \mapsto r) \in \sigma$.
Thus $\sigma \implies \{t=t\sigma\}$.

If the depth of $t$ is greater than one, then $t$ has the form
$f(s_1,\ldots,s_n)$ where $s_1,\dots, s_n\in \Terms$
have depth less than the depth of $t$.
By the inductive hypothesis, for each $i=1$, \dots,~$n$,
we have $\sigma \implies \{s_i = s_i\sigma\}$.
Therefore, applying axiom (\ref{eq-ax:congr}), we have
$\sigma \implies \{t = t\sigma\}$.
\end{proof}

As is common in papers involving equality,
we overload the symbol `$=$' and use it to denote both equality and to
represent syntactic identity.
The context makes it clear what is intended.

\section{The Set-Sharing Domain}
\label{sec: set sharing}

In this section,
we first recall the definition of the $\Sharing$ domain
and present the (classical) abstraction function used for
dealing with idempotent substitutions.
We will then give evidence for the problems arising
when applying this abstraction function to the more general case
of substitutions in rational solved form.

\subsection{The $\Sharing$ Domain}
\label{subsec: sharing domain}

The $\Sharing$ domain is due to Jacobs and Langen \cite{JacobsL89}.
However, we use the definition as presented
in~\cite{BagnaraHZ97b} where the set of variables of interest is 
given explicitly.

\pagebreak[3]
\begin{defn} \summary{(The \emph{set-sharing} lattice.)}
Let
\begin{align*}
  \SG
    &\defeq
      \wpf(\Vars)\setdiff \{ \emptyset \} \\
\intertext{%
  and let
}
  \SH
    &\defeq
      \wp(\SG).
\intertext{%
  The \emph{set-sharing lattice} is given by the set
}
  \SSl
    &\defeq
      \bigl\{\, (\sh, U)
        \bigm| \sh \in \SH, U \in \wpf(\Vars), \forall S \in \sh \itc S \sseq U
      \,\bigr\} \union \{ \bot, \top \},
\end{align*}
which is ordered by `$\leqSSl$' defined as follows,
for each $d, (\sh_1, U_1), (\sh_2, U_2) \in \SSl$:
\begin{align*}
          \bot &\leqSSl d, \\
             d &\leqSSl \top, \\
  (\sh_1, U_1) &\leqSSl (\sh_2, U_2)
    \quad\iff\quad
      (U_1 = U_2)  \land (\sh_1 \sseq \sh_2).
\end{align*}
\end{defn}
It is straightforward to see that every subset of $\SSl$
has a least upper bound with respect to $\leqSSl$.
Hence $\SSl$ is a complete lattice.\footnote{
Notice that the only reason we have $\top \in \SSl$
is in order to turn $\SSl$ into a lattice rather than a CPO.}
The $\lub$ operator over $\SSl$ will be denoted by `$\sqcup$'.

\subsection{The Classical Abstraction Function for $\ISubst$}
\label{subsec:classical-sharing-abstraction}

An element $\sh$ of $\SH$ encodes the sharing information
contained in an idempotent substitution $\sigma$.
Namely, two variables $x$ and $y$ must be in the same set in $\sh$
if some variable occurs in both $x\sigma$ and $y\sigma$.

\begin{defn} \summary{(Classical $\sg$ and abstraction functions.)}
\label{def:langen-occurrence-function}
$\fund{\sg}{\ISubst\times\Vars}{\wpf(\Vars)}$,
called \emph{sharing group function},
is defined,
for each $\sigma \in \ISubst$ and each $v \in \Vars$, by
\[
  \sg(\sigma, v)
    \defeq
        \bigl\{\,
          y \in \Vars
        \bigm|
          v \in \vars(y\sigma)
        \,\bigr\}.
\]
The concrete domain $\wp(\ISubst)$ is related to $\SSl$
by means of the \emph{abstraction function}
$\fund{\alpha_I}{\wp(\ISubst)\times\wpf(\Vars)}{\SSl}$.
For each $\Sigma \in \wp(\ISubst)$ and each $U \in \wpf(\Vars)$,
\begin{align*}
  \alpha_I(\Sigma, U)
    &\defeq
      \bigsqcup_{\sigma \in \Sigma}
        \alpha_I(\sigma, U), \\
\intertext{%
  where $\fund{\alpha_I}{\ISubst\times\wpf(\Vars)}{\SSl}$ is defined,
  for each substitution $\sigma \in \ISubst$ and each $U \in \wpf(\Vars)$, by
}
  \alpha_I(\sigma, U)
    &\defeq
  \Bigl(
    \bigl\{\, \sg(\sigma, v) \inters U \bigm| v \in \Vars \,\bigr\}
      \setdiff \{\emptyset\}, U
  \Bigr).
\end{align*}
\end{defn}
The sharing group function $\sg$ was first defined by 
Jacobs and Langen~\cite{JacobsL89} and used in
their definition of a concretisation function for $\SH$. The function 
$\alpha_I$
corresponds closely to the abstract counterpart of this concretisation
function, but explicitly includes the set of variables of interest as a
separate argument. It is identical to the abstraction function for Sharing
defined by Cortesi and Fil\'e~\cite{CortesiF99}.

In order to provide an intuitive reading of the sharing information encoded
into an abstract element, we should stress that the analysis aims at capturing
\emph{possible} sharing.
The corresponding \emph{definite} information
(e.g., definite groundness or independence)
can be extracted by observing which sharing groups
are \emph{not} in the abstract element.
As an example, if we observe that there is no sharing group containing a
particular variable of $U$, then we can safely conclude that this variable
is definitely ground (namely, it is bound to a term containing no variables).
Similarly, if we observe that two variables never occur together
in the same sharing group,
then we can safely conclude that they are independent
(namely, they are bound to terms that do not share a common variable).
For a more detailed description of the information contained in an element
of $\SSl$, we refer the interested reader
to \cite{BagnaraHZ97b,BagnaraHZ01TCS}.
\begin{exmp}
Assume $U = \{x_1,x_2,x_3,x_4\}$ and let
\[
\sigma =
    \bigl\{
        x_1 \mapsto f(x_2, x_3),
        x_4 \mapsto a
    \bigr\},
\]
so that its abstraction is given by
\[
 \alpha_I(\sigma, U) =
    \Bigl(\bigl\{\{x_1, x_2\}, \{x_1, x_3\}\bigr\}, U\Bigl).
\]
From this abstraction we can safely conclude that
variable $x_4$ is ground and variables $x_2$ and $x_3$ are independent.
\end{exmp}

\subsection{Towards an Abstraction Function for $\RSubst$}
\label{subsec:problems-for-classical-sharing-abstraction}

To help motivate the approach we have taken in adapting the 
classical abstraction function to non-idempotent substitutions,
we now explain some of the problems that arise if we apply $\alpha_I$,
as it is defined on $\ISubst$, to the non-idempotent substitutions in 
$\RSubst$.
Note that these problems are only partially due to allowing for
non-Herbrand substitutions (that is substitutions that are not 
satisfiable in a syntactic equality theory containing the occurs-check axioms).
They are also due to the presence of non-idempotent
but Herbrand substitutions
that may arise because of the potential ``laziness'' of
unification procedures based on the rational solved form.

We use the following substitutions to illustrate the problems,
where it is assumed that the set of variables of interest
is $U = \{x_1,x_2,x_3,x_4\}$. Let
\begin{align*}
\sigma_1 &=
    \bigl\{
        x_1 \mapsto f(x_1)
    \bigr\},\\
\sigma_2 &=
    \bigl\{x_3 \mapsto x_4
    \bigr\},\\
\sigma_3 &=
    \bigl\{
        x_1 \mapsto x_2,
        x_2 \mapsto x_3,
        x_3 \mapsto x_4
    \bigr\},\\
\sigma_4 &=
    \bigl\{
        x_1 \mapsto x_4,
        x_2 \mapsto x_4,
        x_3 \mapsto x_4
    \bigr\}
\end{align*}
so that we have
\begin{align*}
 \alpha_I(\emptyset, U) &= \alpha_I(\sigma_1, U) =
    \Bigl(\bigl\{\{x_1\},\{x_2\},\{x_3\},\{x_4\}\bigr\}, U\Bigl),\\
\alpha_I(\sigma_2, U) &= \alpha_I(\sigma_3, U) =
    \Bigl(\bigl\{\{x_1\}, \{x_2\}, \{x_3, x_4\}\bigr\}, U\Bigl),\\
\alpha_I(\sigma_4, U) &=
    \Bigl(\bigl\{\{x_1, x_2, x_3, x_4\}\bigr\}, U\Bigl).
\end{align*}

The first problem is that the concrete equivalence classes 
induced by the classical abstraction function on $\RSubst$ 
are much coarser than one would expect and hence we have an
unwanted loss of precision.
For example, in all the sets of rational trees that are solutions
for $\sigma_1$, the variable $x_1$ is ground.
However, the computed abstract element
fails to distinguish this situation from that 
resulting from the empty substitution, where all the 
variables are free and un-aliased.
Similarly, we have the same abstract element
for both $\sigma_2$ and $\sigma_3$
although,
$x_1$, $x_2$ and $x_3$ are independent in $\sigma_2$ only.

The second problem is quite the opposite from the first in
 that the abstraction function distinguishes 
between substitutions that are equivalent 
(with respect to any equality theory).
For example, $\sigma_3$ and $\sigma_4$ are equivalent although the 
abstract elements are distinct.
Note that the two problems described here are completely orthogonal
although they can interact and produce more complex situations.

\section{Variable-Idempotence}
\label{sec:Variable-Idempotence}

In this section we define a new class of substitutions
based on the concept of \emph{variable-idempotence}.
Variable-idempotent substitutions are then related to
substitutions in rational solved form by means of
an equivalence preserving rewriting relation.

\subsection{Variable-Idempotent Substitutions}
Recall that, for substitutions, the definition of idempotence
 requires that repeated
applications of a substitution do not change the syntactic structure
of a term.  
However, a sharing abstraction such as $\alpha_I$ is only
interested in the variables and not in the structure that contains them.
Thus, an obvious way to relax the definition of idempotence to
allow for a non-Herbrand substitution is to ignore the structure and
 just require that its repeated
application leaves the set of free variables in a term invariant.

\pagebreak[4]
\begin{defn} 
\label{def:variable-idempotence}
\summary{(Variable-Idempotence.)}  
A substitution $\sigma$ is said to be
\emph{vari\-able-idem\-po\-tent}
if $\sigma \in \RSubst$ and, for each $t \in \Terms$,
\[
  \vars(t\sigma\sigma)\setdiff\dom(\sigma) =
  \vars(t\sigma)\setdiff\dom(\sigma).
\]
The set of all variable-idempotent substitutions is denoted by $\VSubst$.
\end{defn}
Note that, as the condition
$\vars(t\sigma)\setdiff\dom(\sigma) \sseq \vars(t\sigma\sigma)$
is trivial and holds for all substitutions, we have $\sigma \in \VSubst$
if and only if $\sigma \in \RSubst$ and
\begin{equation}
\label{eq:var-idemp-sufficient}
  \vars(t\sigma\sigma)\setdiff\dom(\sigma) \sseq
  \vars(t\sigma).
\end{equation}
Also note that any idempotent substitution is also variable-idempotent,
so that $\ISubst \subset \VSubst \subset \RSubst$.

\begin{exmp}
\label{ex:var-idempotent}
Consider the following substitutions which are all in $\RSubst$.
\begin{alignat*}{2}
\sigma_1 &= \bigl\{x \mapsto f(x)\bigr\} &&\in \VSubst\setdiff \ISubst,\\
\sigma_2 &= \bigl\{x \mapsto f(y), y\mapsto z\bigr\} &&\notin \VSubst,\\
\sigma_3 &= \bigl\{x \mapsto f(z), y\mapsto z\bigr\} &&\in \ISubst,\\
\sigma_4 &= \bigl\{x \mapsto z, y \mapsto f(x,y)\bigr\} &&\notin \VSubst,\\
\sigma_5 &= \bigl\{x \mapsto z, y\mapsto f(z,y)\bigr\} &&\in \VSubst\setdiff \ISubst.
\end{alignat*}
Note that
$\sigma_2$ is equivalent (with respect to any equality theory)
 to the idempotent substitution $\sigma_3$;
and $\sigma_4$
 is equivalent (with respect to any equality theory) to the substitution
 $\sigma_5$ which is variable-idempotent but not idempotent.
\end{exmp}

The next result provides an alternative characterization of variable-idempotence.
\begin{lem}
\label{lem: alt-var-idemp}
Suppose that $\sigma \in \RSubst$.
Then $\sigma \in \VSubst$ if and only if,
for all $(x\mapsto r) \in \sigma$,  
\[
 \vars(r\sigma)\setdiff \dom(\sigma) = \vars(r)\setdiff \dom(\sigma).
\]
\end{lem}
\begin{proof}
Suppose first that $\sigma \in \VSubst$ and that $(x \mapsto r) \in \sigma$.
Then
\[
  \vars(x\sigma\sigma)\setdiff\dom(\sigma)
    = \vars(x\sigma)\setdiff\dom(\sigma)
\]
and hence,
$\vars(r\sigma)\setdiff\dom(\sigma) =
  \vars(r)\setdiff\dom(\sigma)$.

Next, suppose that for all $(x\mapsto r) \in \sigma$,  
$\vars(r\sigma)\setdiff\dom(\sigma) = \vars(r)\setdiff\dom(\sigma)$.
Let $t \in \Terms$.
We will show that
$\vars(t\sigma\sigma)\setdiff\dom(\sigma) =
  \vars(t\sigma)\setdiff\dom(\sigma)$
by induction on the depth of $t$.
If $t$ is a constant or $t \in \Vars\setdiff\dom(\sigma)$, 
then the result follows from the fact that $t\sigma = t$.
If $t \in \dom(\sigma)$, then the result follows from the hypothesis.
Finally, if $t = f(t_1,\ldots,t_n)$, then, by the inductive hypothesis,
$\vars(t_i\sigma\sigma)\setdiff\dom(\sigma) =
  \vars(t_i\sigma)\setdiff\dom(\sigma)$ for $i = 1$, \ldots, $n$.
Therefore we have $\vars(t\sigma\sigma)\setdiff\dom(\sigma) =
  \vars(t\sigma)\setdiff\dom(\sigma)$.
Thus, by Definition~(\ref{def:variable-idempotence}),
as $\sigma \in \RSubst$, $\sigma\in \VSubst$.
\end{proof}

Note that, as a consequence of Lemma~\ref{lem: alt-var-idemp}, 
any substitution consisting of a single binding is
variable-idempotent. Note though that we cannot assume that
every subset of a variable-idempotent substitution is variable-idempotent.
\begin{exmp}
\label{ex:problem-of-subsets}
Let 
\begin{align*}
\sigma_1    &= \{
              x_1 \mapsto x_2,
              x_2 \mapsto g(x_3),
              x_3 \mapsto f(x_3)
           \},\\
\sigma_2  &= \{ 
              x_3 \mapsto f(x_3)
           \},\\
\sigma_3     &= \sigma_1 \setminus \sigma_2
         = \{ 
              x_1 \mapsto x_2,
              x_2 \mapsto g(x_3)
           \}. 
\end{align*}
It can be observed that $\sigma_1,\sigma_2 \in \VSubst$.
Also note that $\sigma_3 \notin \VSubst$, because we have
$x_3 \in \vars(x_1\sigma_3\sigma_3) \setminus \dom(\sigma_3)$
but $x_3 \notin \vars(x_1\sigma_3)\setminus \dom(\sigma_3)$.
\end{exmp}
On the other hand, a variable-idempotent substitution does enjoy the following 
useful property with respect to its subsets.
\begin{lem}
\label{lem:var-idem-subset-property}
If $\sigma \in \VSubst$ and $t \in \Terms$, then, 
for all $\sigma' \sseq \sigma$,
\[
  \vars(t\sigma\sigma')\setdiff\dom(\sigma) =
  \vars(t\sigma)\setdiff\dom(\sigma).
\]
\end{lem}
\begin{proof}
Observe that, since $\sigma' \sseq \sigma$, the relation
\(
  \vars(t\sigma)\setdiff\dom(\sigma) \sseq
  \vars(t\sigma\sigma')
\)
is trivial.

To prove the opposite relation, 
suppose that $y \in \vars(t\sigma\sigma')\setdiff\dom(\sigma)$. 
Then
there exists $x \in \vars(t\sigma)$ such that $y\in \vars(x\sigma')$.
Now, if $x \notin \dom(\sigma')$, then $x = y$ and $y \in \vars(t\sigma)$.
On the other hand, if $x \in \dom(\sigma')$, 
then $x\sigma' = x\sigma$
so that $y \in \vars(t\sigma\sigma)\setdiff\dom(\sigma)$ and hence,
as $\sigma \in \VSubst$, $y\in \vars(t\sigma)$.
\end{proof}
We note that this result depends on the definition of variable-idempotence
ignoring the domain elements of the substitution.
\begin{exmp}
Let 
\[
  \sigma = \bigl\{ x \mapsto f(x,y), y\mapsto a \bigr\}.
\]
Then $\sigma \in \VSubst$ but
\begin{align*}
\vars(x\sigma) &= \{x,y\},\\
\vars(x\sigma\sigma) &= \{x,y\},\\
\vars\bigl(x\sigma\{y\mapsto a\}\bigr) &= \{x\}.
\end{align*}
\end{exmp}

We now state two technical results that will be 
needed later in the paper.
Note that, when proving these results at the end of this section,
we require that the equality theory also satisfies the identity axioms.
They show that equivalent, ordered, variable-idempotent substitutions
have the same domain and bind the domain variables to terms
with the same set of parameter variables.
\begin{lem}
\label{lem: var-idem-equiv-dom}
Suppose that $T$ is a syntactic equality theory,
$\tau,\sigma \in \VSubst$ are ordered and satisfiable in $T$ and
$T \entails \forall(\tau \pimplies \sigma)$.
Then $\dom(\sigma) \sseq \dom(\tau)$.
\end{lem}

\begin{lem}
\label{lem: var-idem-equations}
Suppose that $T$ is a syntactic equality theory,
$\tau,\sigma \in \VSubst$ are
satisfiable in $T$ and
$T \entails \forall(\tau \pimplies \sigma)$.
In addition, suppose $s,t \in \Terms$ are such that
$T \entails \forall\bigl(\tau \pimplies (s = t)\bigr)$.
Then, if $v \in \vars(s)\setdiff\dom(\tau)$,
there exists a variable $z \in \vars(t\sigma)\setdiff \dom(\sigma)$
such that $v \in \vars(z\tau)$.
\end{lem}

\subsection{\cS-transformations}

A useful property of variable-idempotent substitutions is that any substitution
can be transformed to an equivalent (with respect to any equality theory)
variable-idempotent one.

\begin{defn}
\summary{(\cS-transformation.)}
The relation $\reld{\Sstep}{\RSubst}{\RSubst}$,
called \emph{\cS-step}, is defined by
\[
\genfrac{}{}{}{}
  {
    (x \mapsto t) \in \sigma \qquad (y \mapsto s) \in \sigma \qquad x \neq y
  }
  {
    \sigma
      \Sstep
        \bigl(\sigma\setdiff\{y \mapsto s\}\bigr) \union \{y \mapsto s[x/t]\}
  }.
\]
If we have a finite sequence of \cS-steps
$\sigma_1 \Sstep \cdots \Sstep \sigma_n$ mapping $\sigma_1$ to $\sigma_n$, 
then we write $\sigma_1 \Sstepstar \sigma_n$ and say that
$\sigma_1$ can be rewritten, by \cS-transformation, to $\sigma_n$.
\end{defn}

\begin{exmp}
\label{ex:S}
Let
\begin{align*}
  \sigma_0
    &=
      \bigl\{
        x_1 \mapsto f(x_2),
        x_2 \mapsto g(x_3,x_4),
        x_3 \mapsto x_1
      \bigr\}.
\intertext{%
Observe that $\sigma_0$ is not variable-idempotent
since $\vars(x_1\sigma_0)\setdiff \{x_1,x_2,x_3\} = \emptyset$ but 
$\vars(x_1\sigma_0\sigma_0)\setdiff \{x_1,x_2,x_3\} = \{x_4\}$.
By considering all the bindings of the substitution, one at a time,
and applying the corresponding \cS-step to all the other bindings,
we produce a new substitution $\sigma_3$.
}
  \sigma_0
    &=
      \bigl\{
        \underline{x_1 \mapsto f(x_2)},
        x_2 \mapsto g(x_3,x_4),
        x_3 \mapsto x_1
      \bigr\} \\
  \sigma_1
    &=
      \bigl\{
        x_1 \mapsto f(x_2),
        \underline{x_2 \mapsto g(x_3,x_4)},
        x_3 \mapsto f(x_2)
      \bigr\}, \\
  \sigma_2
    &=
      \bigl\{
        x_1 \mapsto f(g(x_3,x_4)),
        x_2 \mapsto g(x_3,x_4),
        \underline{x_3 \mapsto f(g(x_3,x_4))}
     \bigr\}, \\
  \sigma_3
    &=
      \bigl\{
        x_1 \mapsto f(g(f(g(x_3,x_4)),x_4)), \\
        &\phantom{{}=\bigl\{}
        x_2 \mapsto g(f(g(x_3,x_4)),x_4),
        x_3 \mapsto f(g(x_3,x_4))
      \bigr\}.
\end{align*}
Then
\[
  \sigma_0 \Sstepstar \sigma_1 \Sstepstar \sigma_2 \Sstepstar \sigma_3.
\]
Note that $\sigma_0 \iff \sigma_3$ and,
for any $\tau \sseq \sigma_3$, the substitution $\tau$
is variable-idempotent.
In particular, $\sigma_3$ is variable-idempotent.
\end{exmp}

The next two theorems, which are proved at the end of this section, 
show that we need only consider variable-idempotent substitutions.
\pagebreak[3]
\begin{thm}
\label{thm:var-idem-equiv}
Suppose $\sigma \in \RSubst$ and $\sigma \Sstepstar \sigma'$.
Then $\sigma' \in \RSubst$, $\dom(\sigma) = \dom(\sigma')$,
$\vars(\sigma) = \vars(\sigma')$ and,
if $T$ is any equality theory,
then 
$T \entails \forall(\sigma \piff \sigma')$.
\end{thm}

\begin{thm}
\label{thm:var-idem}
Suppose $\sigma \in \RSubst$.
Then there exists  $\sigma' \in \VSubst$ such that
$\sigma \Sstepstar \sigma'$ and, for all $\tau \sseq \sigma'$,
$\tau \in \VSubst$.
\end{thm}
As a consequence of Theorem~\ref{thm:var-idem},
we can transform any substitution in rational solved form
to a substitution for which it and all its subsets are variable-idempotent.
Thus, substitutions such as $\sigma_1$ in Example~\ref{ex:problem-of-subsets}
can be disregarded.
The proof of this theorem formalizes the rewriting process
informally described in Example~\ref{ex:S}.

The following result concerning composition of substitutions
will be needed later.
\begin{lem}
\label{lem: properties-of-composition}
Let $\tau,\sigma \in \VSubst$,
where $\dom(\sigma) \inters \vars(\tau) = \emptyset$.
Then $\tau\circ\sigma$ has the following properties.
\begin{enumerate}
\item
\label{eq:properties-nu-sigma-equiv}
\(
  T \entails  
    \forall\bigl((\tau\circ\sigma) \piff (\tau \union \sigma)\bigr),
\) for any equality theory $T$;
\item
\label{eq:properties-nu-sigma-doms}
\(
  \dom(\tau\circ\sigma) = \dom(\tau \union \sigma);
\)
\item
\label{eq:properties-nu-sigma-videm}
$\tau\circ\sigma \in \VSubst$.
\end{enumerate}
\end{lem}

\subsection{The Abstraction Function for $\VSubst$}

With these results,
it can be seen that we need to consider variable-idempotent
substitutions only.
Moreover, in this case,
one of the causes of the problems outlined
in Section~\ref{subsec:problems-for-classical-sharing-abstraction},
due to the possible ``laziness'' of the unification algorithm,
 is no longer present.
As a consequence, it is now sufficient to address
the potential loss in precision
due to the non-Herbrand substitutions.
The simple solution is to define a new abstraction function for
$\VSubst$
which is the same as that in
Definition~\ref{def:langen-occurrence-function}
but where any sharing group generated by a variable
in the domain of the substitution is disregarded.
This new abstraction function works
for variable-idempotent substitutions
and no longer suffers the drawbacks outlined
in Section~\ref{subsec:problems-for-classical-sharing-abstraction}.

Therefore, at least from a theoretical point of view,
the problem of
defining a sound and precise abstraction function for
arbitrary substitutions in rational solved form
would have been solved.
Given a substitution in $\RSubst$,
we would proceed in two steps:
we first transform it to an equivalent substitution in $\VSubst$
and then compute the corresponding description
by using the modified abstraction function.
However, from a practical point of view,
we need to define an abstraction function
that directly computes the description of a substitution in $\RSubst$
in a single step, thus avoiding the expensive computation
of the intermediate variable-idempotent substitution.
We present such an abstraction function
in Section~\ref{sec:abstracting-in-rsubst}.
 
\subsection{Proofs of Lemmas~\ref{lem: var-idem-equiv-dom}, 
\ref{lem: var-idem-equations} and 
\ref{lem: properties-of-composition} and
Theorems~\ref{thm:var-idem-equiv} and~\ref{thm:var-idem}}

To prove Lemmas~\ref{lem: var-idem-equiv-dom}
and~\ref{lem: var-idem-equations}, 
it is useful to first establish the following two properties of 
variable-idempotent substitutions.

\begin{lem}
\label{lem:L1}
Suppose that $\sigma \in \VSubst$, $r \in \Terms$
and, for all $i \geq 0$, $r\sigma^i \in \Vars$.
Then we have $r\sigma \in \Vars \setdiff \dom(\sigma)$.
\end{lem}
\begin{proof}
As $\sigma$ has no circular subset and 
$\dom(\sigma)$ is finite, 
there exists a $j \geq 1$ such that 
$r\sigma^j = r\sigma^{j+1}$ and hence,
$r\sigma^j  \in \Vars \setminus \dom(\sigma)$.
As $\sigma$ is variable-idempotent, we have
\begin{align*}
\{ r\sigma^j \}
  &= \vars(r\sigma^j) \setminus \dom(\sigma)\\
  &= \vars(r\sigma) \setminus \dom(\sigma)\\
  &= \{ r\sigma \} \setminus \dom(\sigma).
\end{align*}
Hence $r\sigma \in \Vars \setminus \dom(\sigma)$.
\end{proof}

\begin{lem}
\label{lem:L2}
Suppose that $\sigma \in \VSubst$
and $v,r \in \Terms$,
where $v \in \Vars \setdiff \dom(\sigma)$ and,
for any syntactic equality theory $T$,
$T\entails \forall\bigl(\sigma \pimplies\{v=r\}\bigr)$.
Then $v = r\sigma$.
\end{lem}
\begin{proof}
We assume that the congruence and identity axioms hold. 
Let $a_1, a_2\in \GTerms$ have distinct outer-most 
symbols so that,
by the identity axioms, $T \entails a_1 \neq a_2$.
By Lemma~\ref{lem:L1}, either $r\sigma \in \Vars\setdiff\dom(\sigma)$ or,
for some $j \geq 0$, $r\sigma^j \notin \Vars$. 
We consider each case separately.

If, for some $j\geq 0$, $r\sigma^j \notin \Vars$,
then, as $a_1$ and $a_2$ have distinct outer-most symbols, 
there exists an $i \in \{1,2\}$
such that $a_i$ and $r\sigma^j$ have distinct outer-most symbols.
Thus, by the identity axioms, $a_i \neq r\sigma^j$.
Let $\sigma' = \sigma\union \{v = a_i\}$. 
It follows from Lemma~\ref{lem: add-binding} that, 
as $v\notin \dom(\sigma)$ and $\sigma$ is satisfiable, 
$\sigma' \in \RSubst$ and is satisfiable.
By Lemma~\ref{lem: applic} and the congruence axioms, 
$\sigma \implies \{v = r\sigma^j\}$.
However, $\sigma' \implies \sigma$, so that
 $\sigma' \implies \{v = r\sigma^j, v=a_i\}$.
Thus, by the congruence axioms, we have
$\sigma' \implies \{a_i = r\sigma^j\}$,
which is a contradiction.

Suppose then that $r\sigma \in \Vars\setdiff\dom(\sigma)$.
If $v \neq r\sigma$, then it follows from Lemma~\ref{lem: add-binding} that
$\sigma' = \sigma \union \{v = a_1, r\sigma = a_2\} \in \RSubst$ and, 
as $\sigma$ is satisfiable, $\sigma'$ is satisfiable.
By Lemma~\ref{lem: applic} and the congruence axioms, 
$\sigma \implies \{v = r\sigma\}$.
However, $\sigma' \implies \sigma$, so that
$\sigma' \implies \{v = r\sigma, v=a_1, r\sigma = a_2\}$.
Thus, by the congruence axioms, we have 
$\sigma' \implies \{a_1 = a_2\}$,
which is a contradiction. Hence $v = r\sigma$ as required.
\end{proof}

\pagebreak[4]
\begin{proof}[Proof of Lemma~\ref{lem: var-idem-equiv-dom}.]
We assume that the congruence and identity axioms hold. 
To prove the result, we suppose that there exists 
$v \in \dom(\sigma)\setdiff \dom(\tau)$ and 
derive a contradiction.

By hypothesis, $\tau \implies \sigma$. 
Thus, using Lemma~\ref{lem: applic} and the congruence axioms, we have,
for any $i\geq 0$,
$\tau \implies \{v = v\sigma^i\}$.
By Lemma~\ref{lem:L2}, for all $i\geq 0$, $v = v\sigma^i\tau$
so that $v\sigma^i \in \Vars$.
By Lemma~\ref{lem:L1}, $v\sigma \notin \dom(\sigma)$, so that, 
as $\sigma$ is ordered and  $v \in \dom(\sigma)$, $v\sigma < v$.
In particular, $v\sigma \neq v$,
so that as $v\sigma\tau = v$ and $\tau$ is ordered,
we would have $v < v\sigma$, which is a contradiction.
\end{proof}

\begin{proof}[Proof of Lemma~\ref{lem: var-idem-equations}.]
We assume that the congruence and identity axioms hold.
Note that, by the hypothesis,
$\tau \implies \sigma$ and $\tau \implies \{s = t\}$ so that,
using Lemma~\ref{lem: applic} and the congruence axioms, we have
$\tau \implies \{s = t\sigma^j\}$ and 
$\tau \implies \{t\sigma\tau^k = s\}$,
for all $j,k \geq 0$.

Let $v \in \vars(s)\setdiff\dom(\tau)$.
We prove, by induction on the depth $d$ of $s$, 
that there exists $z \in \vars(t\sigma)\setdiff \dom(\sigma)$
such that $v \in \vars(z\tau)$.
The base case is when $d=1$ so that $s = v$.
Now, for each $j \geq 0$, $\tau \implies \{v = t\sigma^j\}$
and hence, by  Lemma~\ref{lem:L2} (as $v \notin \dom(\tau)$),
$v = t\sigma^j\tau$.
As a consequence, $t\sigma^j \in \Vars$ for all $j \geq 0$ and 
$v = t\sigma\tau$.
By Lemma~\ref{lem:L1}, $t\sigma \in \Vars\setdiff \dom(\sigma)$.
Thus, we define $z = t\sigma$.

For the inductive step, we assume that
$d > 1$ so that,
for some $n \geq 1$, we have $s = f(s_1,\ldots,s_n)$ and,
for some $i \in \{1, \ldots, n\}$, $v\in \vars(s_i)$ and $s_i$ has depth $d-1$.
By  Lemma~\ref{lem:L1}, either $t\sigma \in \Vars\setdiff \dom(\sigma)$ or 
there exists a $j \geq 0$ such that $t\sigma^j \notin \Vars$.

First, suppose that $t\sigma \in \Vars\setdiff \dom(\sigma)$.
Now, $\tau \implies \{t\sigma\tau = s\}$ so that,
as $s\tau \notin \Vars$, by  Lemma~\ref{lem:L2},
we have $t\sigma\tau \notin \Vars\setdiff\dom(\tau)$.
Thus, by  Lemma~\ref{lem:L1},
 there exists $k > 1$ such that $t\sigma\tau^k \notin \Vars$.
Then, using the identity axioms, we have
$t\sigma\tau^k = f(r_1,\ldots,r_n)$ and
$\tau \implies \{s_i = r_i\}$.
By the inductive hypothesis (letting $\sigma$ be the empty substitution),
we have $v \in \vars(r_i\tau)$.
However, $\vars(r_i) \sseq \vars(t\sigma\tau^k)$ so that
$v \in \vars(t\sigma\tau^{k+1})$.
As $\tau \in \VSubst$ and $v \notin \dom(\tau)$,
$v \in \vars(t\sigma\tau)$.
Thus, in this case, let $z = t\sigma$.

Secondly, suppose that there exists a $j \geq 0$ such that 
$t\sigma^j \notin \Vars$.
Then, as $\tau \implies \{s = t\sigma^j\}$,
it follows from the identity axioms
that $t\sigma^j = f(t_1,\ldots,t_n)$ and
$\tau \implies \{s_i = t_i\}$.
By the inductive hypothesis,
there exists $z\in \vars(t_i\sigma)\setdiff\dom(\sigma)$ such that $v \in \vars(z\tau)$.
However, $\vars(t_i\sigma) \sseq \vars(t\sigma^{j+1})$ so that we must have
$z \in \vars(t\sigma^{j+1})\setdiff\dom(\sigma)$.
As $\sigma \in \VSubst$,
$z \in \vars(t\sigma)\setdiff\dom(\sigma)$ as required.
\end{proof}

To prove Theorem~\ref{thm:var-idem-equiv},
we need to show that the result holds for a single \cS-step.

\begin{lem}
\label{lem:S-trans-equiv}
Let $T$ be an equality theory
and suppose that $\sigma\in \RSubst$ and
$\sigma \Sstep \sigma'$.
Then $\sigma' \in \RSubst$,
 $\dom(\sigma) = \dom(\sigma')$, $\vars(\sigma) = \vars(\sigma')$,
and
$T \entails \forall(\sigma \piff \sigma')$.
\end{lem}
\begin{proof}
Since $\sigma \Sstep \sigma'$,
there exists $x,y\in \dom(\sigma)$ with $x \neq y$ such that
$(x \mapsto t), (y \mapsto s) \in \sigma$ and
\(
  \sigma'
    = \bigl(\sigma\setdiff\{y \mapsto s\}\bigr)
      \union \bigl\{y \mapsto s[x/t]\bigr\}
\).
If $x \notin \vars(s)$,  $\sigma = \sigma'$ and the result is trivial.
Suppose now that $x \in \vars(s)$.
We define
\begin{align}
\notag
\sigma_0 &\defeq \sigma \setdiff \{x=t,y=s\}.\\
\intertext{%
Hence, as it is assumed that $x\neq y$,
}
\label{eq:lem:S-trans-equiv:sigma}
\sigma &= \sigma_0 \union \{x \mapsto t, y \mapsto s\}, \\
\label{eq:lem:S-trans-equiv:sigma'}
\sigma' &= \sigma_0 \union \{x \mapsto t, y \mapsto s[x/t]\}.
\end{align}

We first show that $\sigma' \in \RSubst$ and
$\dom(\sigma) = \dom(\sigma')$.
If $s \notin \Vars$, then $s[x/t] \notin \Vars$ so that
$\dom(\sigma) = \dom(\sigma')$.
Also, as
$\sigma$ has no circular subset,
$\sigma'$ has no circular subset and $\sigma' \in \RSubst$.
If $s \in \Vars$, then $s=x$ and $s[x/t] = t$.
Thus,
as $\sigma = \sigma_0 \union \{x \mapsto t, y\mapsto x\}$ 
has no circular subset,  $t \neq y$ so that
$\dom(\sigma) = \dom(\sigma')$. 
Moreover, neither $\sigma_0 \union \{x \mapsto t\}$ nor
$\sigma_0 \union \{y \mapsto t\}$ have circular subsets. 
Hence $\sigma'$ has no circular subset.
Thus $\sigma' \in \RSubst$.

Now, since
\[
  \bigl(\vars(s) \union \vars(t)\bigr) \setdiff \dom(\sigma)
    = \vars\bigl(s[x/t] \union \vars(t)\bigr) \setdiff \dom(\sigma),
\]
it follows that $\vars(\sigma) = \vars(\sigma')$.

Therefore, it remains to show that, for any equality theory $T$,
$T \entails \forall(\sigma \piff \sigma')$.
To do this, we assume that
the congruence axioms hold, and show that $\sigma \iff \sigma'$.
By Lemma~\ref{lem: applic}, we have
\begin{align*}
  \{x=t\} &\implies  \{s = s[x/t]\}.\\
\intertext{%
Thus, using the congruence axiom~(\ref{eq-ax:trans}), we have
}
 \{ x=t, y=s \} &\implies \bigl\{ x=t, y=s, s = s[x/t] \bigr\}\\
          &\implies \bigl\{ x=t, y=s[x/t] \bigr\}.\\
\intertext{%
Similarly, using
congruence axioms~(\ref{eq-ax:sym}) and~(\ref{eq-ax:trans}), we have
}
 \bigl\{ x=t, y=s[x/t] \bigr\}
          &\implies \bigl\{ x=t, y=s[x/t], s = s[x/t] \bigr\}\\
          &\implies \{ x=t, y=s \}.\\
\intertext{%
Thus
}
\{x=t, y=s\} &\iff \bigl\{ x=t, y=s[x/t] \bigr\}.
\end{align*}
It therefore follows from 
(\ref{eq:lem:S-trans-equiv:sigma}) and 
(\ref{eq:lem:S-trans-equiv:sigma'}) that 
$\sigma \iff \sigma'$.
\end{proof}

The condition $x \neq y$ in the proof of
Lemma~\ref{lem:S-trans-equiv} is necessary.
For example, suppose
$\sigma = \bigl\{x\mapsto f(x)\bigr\}$ and
$\sigma' = \bigl\{x\mapsto f(f(x))\bigr\}$. Then we do not have
$\sigma' \implies \sigma$.
Note however that this implication will hold
as soon as we enrich the equality theory $T$ with either
the occurs-check axioms or the uniqueness axioms of the
rational trees' theory.

\begin{proof}[Proof of Theorem~\ref{thm:var-idem-equiv}.]
The proof is by induction on the length of the sequence of \cS-steps 
transforming $\sigma$ to $\sigma'$.
The base case is the empty sequence.
For the inductive step, the sequence has length $n>0$ and 
there exists $\sigma_1$ such that 
$\sigma \Sstep \sigma_1 \Sstepstar \sigma'$ and 
$\sigma_1 \Sstepstar \sigma'$ has length $n-1$.
By Lemma~\ref{lem:S-trans-equiv},
$\sigma_1 \in \RSubst$, 
$\dom(\sigma) = \dom(\sigma_1)$,
$\vars(\sigma) = \vars(\sigma_1)$ and
$T \entails \forall(\sigma \piff \sigma_1)$.
By the inductive hypothesis,
$\sigma' \in \RSubst$, 
$\dom(\sigma_1) = \dom(\sigma')$,
$\vars(\sigma_1) = \vars(\sigma')$ and
$T \entails \forall(\sigma_1 \piff \sigma')$.
Hence we have
$\dom(\sigma) = \dom(\sigma')$,
$\vars(\sigma) = \vars(\sigma')$, and
$T \entails \forall(\sigma \piff \sigma')$.
\end{proof}

\begin{proof}[Proof of Theorem~\ref{thm:var-idem}.]
To prove the theorem,
we construct an \cS-transformation
and show that the resulting substitution has the required properties.

Suppose that $\{x_1,\ldots,x_n\} = \dom(\sigma)$, 
$\sigma_0 = \sigma$ and, for each $j=0$, \dots,~$n$,
\[
  \sigma_j = \{ x_1 \mapsto t_{1,j}, \ldots, x_n \mapsto t_{n,j} \},
\]
where,
if $j>0$, $t_{j,j} = t_{j,j-1}$ and, for each $i=1$, \dots,~$n$
with $i \neq j$, we have $t_{i,j} = t_{i,j-1}[x_j/t_{j,j}]$.

It follows from the definition of $\sigma_j$ that,
for  $j = 1$, \dots,~$n$  ,
$\sigma_j$ can be obtained from
$\sigma_{j-1}$ by two sequences of
 \cS-steps of lengths $j-1$ and $n -j +1$:
\[
  \sigma_{j-1} = \sigma_{j-1}^0
   \Sstep
    \cdots
   \Sstep
  \sigma_{j-1}^{j-1} =
  \sigma_{j-1}^j
   \Sstep
    \cdots
   \Sstep
   \sigma_{j-1}^n = \sigma_j,
\]
where, for  $i = 1$, \dots,~$n$ with $i\neq j$,
\begin{align*}
  \sigma_{j-1}^i
   &= \bigl(\sigma_{j-1}^{i-1}\setdiff \{x_i \mapsto t_{i,j-1}\}\bigr)
    \union
    \bigl\{x_i \mapsto t_{i,j-1}[x_j/t_{j,j}]\bigr\} \\
   &= \{x_1 \mapsto t_{1,j},\,         \ldots,\, x_i\mapsto t_{i,j},\,
         x_{i+1} \mapsto t_{i+1,j-1},\, \ldots,\, x_n \mapsto t_{n,j-1}\}.
\end{align*}
Hence, by Theorem~\ref{thm:var-idem-equiv},
$\sigma_1, \ldots, \sigma_n \in \RSubst$.

We next show, by induction on $j$, with $0\leq j\leq n$,
that, for each $i=1$, \dots,~$n$ and each $h=1$, \dots,~$j$,
we have $\vars(t_{i,j}) = \vars\bigl(t_{i,j}[x_h/t_{h,j}]\bigr)$.

For the base case when $j=0$ there is nothing to prove.
Suppose, therefore, that $1 \leq j \leq n$ and that,
for each $i=1$, \dots,~$n$ and $h=1$, \dots,~$j-1$,
\begin{equation*}
\vars(t_{i,j-1}) = \vars\bigl(t_{i,j-1}[x_h/t_{h,j-1}]\bigr).
\end{equation*}

Now by the definition of $t_{k,j}$ where $1 \leq k\leq n$, $k\neq j$,
we have
\begin{align}
\label{var-idem-tkj}
\vars(t_{k,j}) &= \vars\bigl(t_{k,j-1}[x_j/t_{j,j}]\bigr).\\
\intertext{%
Also, since a substitution consisting of a single binding is
variable-idempotent,}
\notag
\vars(t_{j,j}) &= \vars\bigl(t_{j,j}[x_j/t_{j,j}]\bigr)\\
\intertext{%
so that, as $t_{j,j} = t_{j,j-1}$,}
\label{var-idem-tjj}
\vars(t_{j,j}) &= \vars\bigl(t_{j,j-1}[x_j/t_{j,j}]\bigr).\\
\intertext{%
Thus, by~(\ref{var-idem-tkj}) and (\ref{var-idem-tjj}), 
for all $k$ such that $1 \leq k\leq n$,
we have
}
\label{var-idem-tkj-tjj}
\vars(t_{k,j}) &= \vars\bigl(t_{k,j-1}[x_j/t_{j,j}]\bigr).
\end{align}
Therefore, for each $i=1$, \dots,~$n$ and $h=1$, \dots,~$j$, using
(\ref{var-idem-tkj-tjj}) and the inductive hypothesis, we have
\begin{align*}
\vars\bigl(t_{i,j}[x_h/t_{h,j}]\bigr)
&= \vars\Bigr(
   t_{i,j-1}[x_j/t_{j,j}]\bigl[x_h/t_{h,j-1}[x_j/t_{j,j}]\bigr]
        \Bigr)\\
&= \vars\bigl(t_{i,j-1}[x_h/t_{h,j-1}][x_j/t_{j,j}]\bigr)\\
&= \vars\bigl(t_{i,j-1}[x_j/t_{j,j}]\bigr)\\
&= \vars(t_{i,j}).
\end{align*}
Letting $j=n$ we obtain, for each $i,h=1$, \dots,~$n$,
\begin{align*}
\vars\bigl(t_{i,n}[x_h/t_{h,n}]\bigr) &= \vars(t_{i,n}).\\
\intertext{%
Therefore,
for all $\tau \sseq \sigma_n$ and each $i=1$, \dots,~$n$, 
}
\vars(t_{i,n}\tau) &= \vars(t_{i,n}). 
\end{align*}
Thus, by Lemma~\ref{lem: alt-var-idemp},
for all $\tau \sseq \sigma_n$, 
$\tau \in \VSubst$.
The result follows by taking $\sigma' = \sigma_n$.
\end{proof}

\begin{proof}[Proof of Lemma~\ref{lem: properties-of-composition}.]
Since $\tau$, $\sigma \in \VSubst$ and
$\dom(\sigma) \inters \vars(\tau) = \emptyset$,
we have that $(\tau \union \sigma) \in \RSubst$.
It follows from Eq.~(\ref{eq:alt-compose}) that 
$\tau\circ\sigma$ can be obtained from $(\tau \union \sigma)$ 
by a sequence of \cS-steps so that,
by Theorem~\ref{thm:var-idem-equiv},
we have Properties~\ref{eq:properties-nu-sigma-equiv} and
\ref{eq:properties-nu-sigma-doms}.

To prove Property \ref{eq:properties-nu-sigma-videm},
we suppose that, for some $v\in \dom(\tau \circ \sigma)$, there exist
$w \in \vars(v\sigma)$,
$x \in \vars(w\tau)$ and
$y\in \vars(x\sigma)$ such that
$z \in \vars(y\tau)\setdiff\dom(\tau\circ\sigma)$.
We need to prove that $z \in \vars(v\sigma\tau)$.

It follows from Property~\ref{eq:properties-nu-sigma-doms},
that $z\notin\dom(\sigma)$ and $z\notin\dom(\tau)$.
Suppose first that $x\notin\dom(\sigma)$.
Then $y=x$ and hence
$z\in\vars(v\sigma\tau\tau)$.
Therefore, as $\tau \in \VSubst$ and $z \notin \dom(\tau)$,
we can conclude $z \in \vars(v\sigma\tau)$.
Thus, we now assume that $x \in \dom(\sigma)$.
As $\dom(\sigma) \inters \vars(\tau) = \emptyset$,
we have $x \notin \vars(\tau)$,
so that $x=w$ and hence, $y \in \vars(v\sigma\sigma)$.
If $y \notin \dom(\tau)$ we have $y=z$,
so that $y \notin \dom(\sigma)$.
On the other hand, if $y \in \dom(\tau)$ then,
by the hypothesis, $y \notin \dom(\sigma)$.
Thus, in both cases, as $\sigma \in \VSubst$,
we obtain $y \in \vars(v\sigma)$
and hence $z \in \vars(v\sigma\tau)$.
It follows, using Eq.~(\ref{eq:var-idemp-sufficient}),
that Property~\ref{eq:properties-nu-sigma-videm} holds.
\end{proof}    

\section{The Abstraction Function for $\RSubst$}
\label{sec:abstracting-in-rsubst}

In this section we define a new abstraction function mapping
arbitrary substitutions in rational solved form into their
abstract descriptions.
This abstraction function is based on a new definition
for the notion of \emph{occurrence}.
The new occurrence operator $\occ$ is defined on $\RSubst$
so that it does not require the explicit computation of
intermediate variable-idempotent substitutions.
To this end, it is given as the fixed point of a sequence
of \emph{occurrence functions}.
The $\occ$ operator generalises the $\sg$ operator,
defined for $\ISubst$, coinciding with it when
applied to idempotent substitutions.

\pagebreak[4]
\begin{defn} \summary{(Occurrence functions.)}
\label{def:occurrence-functs}
For each $n \in \Nset$,
$\fund{\occ_n}{\RSubst\times\Vars}{\wpf(\Vars)}$,
called \emph{occurrence function},
is defined,
for each $\sigma \in \RSubst$ and each $v \in \Vars$, by
\begin{align*}
  \occ_0(\sigma, v)
    &\defeq
        \{v\}\setdiff \dom(\sigma) \\
\intertext{%
  and, for $n > 0$, by
}
  \occ_n(\sigma, v)
    &\defeq
      \bigl\{\,
        y \in \Vars
      \bigm|
        \vars(y\sigma)\inters \occ_{n-1}(\sigma,v) \neq \emptyset
      \,\bigr\}.
\end{align*}
\end{defn}
The following monotonicity property for $\occ_n$ is proved at the
end of this section.
\begin{lem}
\label{lem:occ-functions-fixed}
If $n>0$, then, for each $\sigma \in \RSubst$ and each $v \in \Vars$,
\[
\occ_{n-1}(\sigma,v) \sseq \occ_n(\sigma,v).
\]
\end{lem}
Note that, by considering the substitution 
$\{u\mapsto v, v\mapsto w\}$, 
it can be seen that, if we had not excluded the domain variables
in the definition of $\occ_0$, then this monotonicity property would
not have held.

For any $n$, the set $\occ_n(\sigma,v)$ is restricted to the set
$\{v\} \union \vars(\sigma)$.
Thus, it follows from Lemma~\ref{lem:occ-functions-fixed},
that there is an $\ell = \ell(\sigma, v) \in \Nset$ such that
$\occ_\ell(\sigma, v) = \occ_{n}(\sigma,v)$ for all $n \geq \ell$.
\begin{defn} \summary{(Occurrence operator.)}
\label{def:occurrence-op}
For each $\sigma \in \RSubst$ and $v \in \Vars$,
the \emph{occurrence operator}
$\fund{\occ}{\RSubst\times\Vars}{\wpf(\Vars)}$
is given by
\[
 \occ(\sigma,v) \defeq \occ_\ell(\sigma,v)
\]
where  $\ell \in \Nset$ is such that
$\occ_\ell(\sigma, v) = \occ_{n}(\sigma,v)$ for all $n \geq \ell$.
\end{defn}
Note that, by combining Definitions~\ref{def:occurrence-functs}
and~\ref{def:occurrence-op}, we obtain
\begin{equation}
\label{eq:property-of-occ}
  \occ(\sigma,v)
    =
      \bigl\{\,
        y \in \Vars
      \bigm|
        \vars(y\sigma) \inters \occ(\sigma, v) \neq \emptyset
      \,\bigr\}.
\end{equation}

The following simpler characterisations for $\occ$
can be used when the variable is in the domain of the substitution,
the substitution is variable-idempotent
or the substitution is idempotent.

\begin{lem}
\label{lem:var-idem-dom-charac}
If $\sigma \in \RSubst$ and $v \in \dom(\sigma)$,
 then $\occ(\sigma,v)= \emptyset$.
\end{lem}

\begin{lem}
\label{lem:var-idem-occ-charac}
If $\sigma \in \VSubst$ then, for each $v \in \Vars$,
\begin{align*}
\occ(\sigma,v) &= \occ_1(\sigma,v)\\
                   &=\bigl\{\,
        y \in \Vars
      \bigm|
        v \in \vars(y\sigma) \setdiff \dom(\sigma)
      \,\bigr\}.
\end{align*}
\end{lem}

\begin{lem}
\label{lem:occ-generalises-sg}
If $\sigma \in \ISubst$ and $v \in \Vars$ then
$\occ(\sigma,v) = \sg(\sigma, v)$.
\end{lem}
These results are proved at the end of this section.

\begin{exmp}
\label{ex:occ}
Consider again \textup{Example~\ref{ex:S}}.
Then, for all $i\geq 0$, $\dom(\sigma_i) = \{x_1,x_2,x_3\}$ so that
\[
\occ(\sigma_i, x_1) 
= \occ(\sigma_i, x_2) 
= \occ(\sigma_i, x_3) 
= \emptyset.
\]
However,
\begin{align*}
\occ_0(\sigma_0, x_4) &= \{x_4\}, \\
\occ_1(\sigma_0, x_4) &= \{x_2,x_4\}, \\
\occ_2(\sigma_0, x_4) &= \{x_1,x_2,x_4\}, \\
\occ_3(\sigma_0, x_4) &= \{x_1,x_2,x_3,x_4\}
 = \occ(\sigma_0, x_4). \\
\intertext{%
Also, note that
}
\occ_1(\sigma_3, x_4) &= \{x_1,x_2,x_3,x_4\}
                    = \occ(\sigma_3, x_4).
\end{align*}
\end{exmp}

The definition of abstraction is based on the occurrence operator, $\occ$.

\begin{defn} \summary{(Abstraction.)}
The concrete domain $\wp(\RSubst) $ is related to $\SSl$
by means of the \emph{abstraction function}
$\fund{\alpha}{\wp(\RSubst)\times\wpf(\Vars)}{\SSl}$.
For each $\Sigma \in \wp(\RSubst)$ and each $U \in \wpf(\Vars)$,
\begin{align*}
  \alpha(\Sigma, U)
    &\defeq
      \bigsqcup_{\sigma \in \Sigma}
        \alpha(\sigma, U) \\
\intertext{%
  where $\fund{\alpha}{\RSubst\times\wpf(\Vars)}{\SSl}$ is defined,
  for each substitution $\sigma \in \RSubst$ and each $U \in \wpf(\Vars)$, by
}
  \alpha(\sigma, U)
    &\defeq
  \Bigl(
    \bigl\{\, \occ(\sigma, v)\inters U \bigm| v\in \Vars \,\bigr\}
      \setdiff \{\emptyset\}, U
  \Bigr).
\end{align*}
\end{defn}

\begin{exmp}
\label{ex:alpha}
Let us consider \textup{Examples~\ref{ex:S}} and~\textup{\ref{ex:occ}}
once more.
Then, assuming $U = \{x_1,x_2,x_3,x_4\}$,
\[
 \alpha(\sigma_0, U) = \Bigl(\bigl\{\occ(\sigma_0,x_4)\bigr\},U\Bigr) =
    \Bigl(\bigl\{\{x_1,x_2,x_3,x_4\}\bigr\}, U\Bigl).
\]
As a second example, consider the substitution
\[
 \sigma = \bigl\{
             x_1 \mapsto f(x_1),
             x_2 \mapsto x_1,
             x_3 \mapsto x_1,
             x_4 \mapsto x_2
          \bigr\}.
\]
Then
\[
\occ(\sigma,x_1) 
   = \occ(\sigma,x_2)
   = \occ(\sigma,x_3)
   = \occ(\sigma,x_4)
   = \emptyset
\]
so that, if we again assume $U = \{x_1,x_2,x_3,x_4\}$,
\[
 \alpha\bigl(\sigma, U\bigr) = \bigl(\emptyset, U\bigr).
\]
\end{exmp}

Any substitution in rational solved form is equivalent,
with respect to any equality theory,
to a variable-idempotent substitution
having the same abstraction.

\pagebreak[4]
\begin{thm}
\label{thm:equiv-ordered-vsubst}
If $T$ is an equality theory and
$\sigma \in \RSubst$ is satisfiable in $T$, then there exists a
substitution $\sigma' \in \VSubst$
such that $\tau \in \VSubst$, for all $\tau \sseq \sigma'$, 
$T \entails \forall(\sigma \piff \sigma')$,
$\vars(\sigma) = \vars(\sigma')$ and
$\alpha(\sigma, U) = \alpha(\sigma', U)$, for any $U \in \wpf(\Vars)$.
\end{thm}

Equivalent substitutions in rational solved form
have the same abstraction.
We note that this property is essential
for the implementation of the $\SSl$ domain.
\begin{thm}
\label{thm:equiv-subst}
If $T$ is a syntactic equality theory and
$\sigma, \sigma' \in \RSubst$ are satisfiable in $T$
and such that $T \entails \forall(\sigma \piff \sigma')$, then
$\alpha(\sigma, U) = \alpha(\sigma', U)$, for any $U \in \wpf(\Vars)$.
\end{thm}

\subsection{Proofs of
Lemmas~\ref{lem:occ-functions-fixed},
\ref{lem:var-idem-dom-charac}, 
\ref{lem:var-idem-occ-charac} and
\ref{lem:occ-generalises-sg}
and
Theorems~\ref{thm:equiv-ordered-vsubst}
and~\ref{thm:equiv-subst}}

\begin{proof}[Proof of Lemma~\ref{lem:occ-functions-fixed}.]
The proof is by induction on $n$.
For the base case (when $n=1$), if $occ_0(\sigma,v) \neq \emptyset$, then
$v\notin\dom(\sigma)$ and $occ_0(\sigma,v) = \{v\}$.
Thus, $v = v\sigma$ so that,
by Definition~\ref{def:occurrence-functs}, $v \in \occ_1(\sigma,v)$.
Suppose $n>1$.
Then, if $y \in \occ_{n-1}(\sigma,v)$, we have,
by Definition~\ref{def:occurrence-functs},
$\vars(y\sigma) \inters \occ_{n-2}(\sigma,v) \neq \emptyset$.
By the induction hypothesis,
\[
  \occ_{n-2}(\sigma,v) \sseq \occ_{n-1}(\sigma,v)
\]
so that
$\vars(y\sigma) \inters \occ_{n-1}(\sigma,v) \neq \emptyset$
and thus $y \in \occ_n(\sigma,v)$.
\end{proof}

\begin{proof}[Proof of Lemma~\ref{lem:var-idem-dom-charac}.]
By Definition~\ref{def:occurrence-functs},
$\occ_0(\sigma,v)= \emptyset$ and, for all $n>0$,
we have $\occ_n(\sigma,v)= \emptyset$
if $\occ_{n-1}(\sigma,v)= \emptyset$.
Thus, $\occ_n(\sigma,v)= \emptyset$, for all $n \geq 0$,
so that, by Definition~\ref{def:occurrence-op}, $\occ(\sigma,v) = \emptyset$.
\end{proof}

\begin{proof*}[Proof of Lemma~\ref{lem:var-idem-occ-charac}.]
Suppose first that $v\in \dom(\sigma)$.
Then
\begin{align*}
 \bigl\{\,
         y \in \Vars
       \bigm|
         v \in \vars(y\sigma) \setdiff \dom(\sigma)
       \,\bigr\}
    &= \emptyset.
\end{align*}
Also, by Lemma~\ref{lem:var-idem-dom-charac},
$\occ_1(\sigma,v) = \occ(\sigma,v) = \emptyset$.

Suppose next that $v\notin \dom(\sigma)$.
It follows from Definition~\ref{def:occurrence-functs}, that
\begin{align*}
  \occ_0(\sigma,v)
    &=
      \{v\}, \\
  \occ_1(\sigma,v)
    &=
      \bigl\{\,
        y \in \Vars
      \bigm|
        \vars(y\sigma) \inters \{v\} \neq \emptyset
      \,\bigr\} \\
    &=
      \bigl\{\,
        y \in \Vars
      \bigm|
        v \in \vars(y\sigma)
      \,\bigr\}, \\
\intertext{%
  and
}
  \occ_2(\sigma,v)
    &=
      \Bigl\{\,
        y \in \Vars
      \Bigm|
        \vars(y\sigma) \inters
           \bigl\{\, y_1 \in \Vars \mid v \in \vars(y_1\sigma) \,\bigr\}
             \neq \emptyset
      \,\Bigr\} \\
    &=
      \bigl\{\,
        y \in \Vars
      \bigm|
        v \in \vars(y\sigma^2)
      \,\bigr\}.
\end{align*}
However, as $\sigma \in \VSubst$, we have
$\vars(y\sigma)\setdiff\dom(\sigma) = \vars(y\sigma^2)\setdiff\dom(\sigma)$.
Thus, as $v\notin \dom(\sigma)$,
$\occ_1(\sigma,v) = \occ_2(\sigma,v)$
and hence, by Definition~\ref{def:occurrence-functs}, we have also
$\occ_n(\sigma,v) = \occ_1(\sigma,v)$, for all $n \geq 1$.
Therefore, by Definition~\ref{def:occurrence-op},
\[
  \occ(\sigma,v)
    =
      \occ_1(\sigma,v)
        =
          \bigl\{\,
            y \in \Vars
          \bigm|
            v \in \vars(y\sigma)
          \,\bigr\}.
\mathproofbox
\]
\end{proof*}

\begin{proof}[Proof of Lemma~\ref{lem:occ-generalises-sg}.]
As $\sigma \in \ISubst$ we have, for all $y \in \Vars$,
$\vars(y\sigma) \setminus \dom(\sigma) = \vars(y\sigma)$.
Also, as $\sigma \in \VSubst$,
we can apply Lemma~\ref{lem:var-idem-occ-charac}
so that
\begin{align*}
  \occ(\sigma,v)
    &=
      \bigl\{\,
        y \in \Vars
      \bigm|
        v \in \vars(y\sigma) \setminus \dom(\sigma)
      \,\bigr\} \\
    &=
      \bigl\{\,
        y \in \Vars
      \bigm|
        v \in \vars(y\sigma)
      \,\bigr\} \\
    &=
      \sg(\sigma,v).
\end{align*}
\end{proof}

To prove Theorem~\ref{thm:equiv-ordered-vsubst}, we need to show that
the abstraction function $\alpha$
is invariant with respect to \cS-transformation.
\begin{lem}
\label{lem: S-equiv subst}
Let
$\sigma,\sigma' \in \RSubst$
where
$\sigma \Sstepstar \sigma'$ and
$U \in \wpf(\Vars)$.
Then $\alpha(\sigma, U) = \alpha(\sigma', U)$.
\end{lem}

\begin{proof}
Suppose first that $\sigma \Sstep \sigma'$.
Thus we assume that $(x \mapsto t), (y \mapsto s) \in \sigma$ ,
where $x \neq y$, and that
\begin{equation}
\label{eq: S-equiv-subst sigma'}
\sigma' = \bigl(\sigma\setdiff \{y\mapsto s\}\bigr) \union
            \bigl\{y\mapsto s[x/t]\bigr\}.
\end{equation}
Suppose $v \in \Vars$.
Then we show that $\occ(\sigma, v) = \occ(\sigma', v)$.

If $x\notin \vars(s)$, then $\sigma' = \sigma$ and there is nothing to
prove.
Also, if $v \in \dom(\sigma)$ then, by Theorem~\ref{thm:var-idem-equiv},
$v \in \dom(\sigma')$ so that\, by Lemma~\ref{lem:var-idem-dom-charac},
$\occ(\sigma,v) = \occ(\sigma',v) = \emptyset$.

We now assume that $x\in \vars(s)$ and
 $v = v\sigma = v\sigma'$. We first prove that,
for each $m \geq 0$,
\begin{equation}
\label{eq: S-equiv-subst occmsigma-in-occbangsigma'}
  \occ_m(\sigma, v) \sseq \occ(\sigma', v).
\end{equation}
The proof is by induction on $m$.
By Definition~\ref{def:occurrence-functs},
we have that
\[
  \occ_0(\sigma, v) = \occ_0(\sigma', v) = \{v\},
\]
so that (\ref{eq: S-equiv-subst occmsigma-in-occbangsigma'}) holds for
$m=0$.
Suppose then that $m>0$ and that
\(
  v_m \in \occ_m(\sigma, v).
\)
Then,
to prove  (\ref{eq: S-equiv-subst occmsigma-in-occbangsigma'}),
we must show that
\(
  v_m \in \occ(\sigma', v).
\)
By Definition~\ref{def:occurrence-functs},
there exists
\begin{equation}
\label{eq: S-equiv-subst vm-1-in-sigma}
 v_{m-1}\in \vars (v_m\sigma) \inters \occ_{m-1}(\sigma, v).
\end{equation}
Hence, by the inductive hypothesis, $v_{m-1} \in \occ(\sigma', v)$.
If $v_{m-1}\in \vars (v_m\sigma')$,
then, by Eq.~(\ref{eq:property-of-occ}), $v_m \in \occ(\sigma', v)$.
Suppose now that $v_{m-1}\notin \vars (v_m\sigma')$.
Since, by~(\ref{eq: S-equiv-subst vm-1-in-sigma}),
we have that $v_{m-1} \in \vars(v_m\sigma)$,
it follows, using (\ref{eq: S-equiv-subst sigma'}), that
$v_m = y$ and $v_{m-1} = x$.
However, by assumption, $v \notin \dom(\sigma)$, so that $x \neq v$ and $m>1$.
Thus, by Definition~\ref{def:occurrence-functs},
there exists
\begin{equation}
\label{eq: S-equiv-subst vm-2-in-sigma}
  v_{m-2}\in \vars (x\sigma) \inters \occ_{m-2}(\sigma,v).
\end{equation}
However, $x\sigma = t$ and $x\in \vars(s)$ so that,
by~(\ref{eq: S-equiv-subst vm-2-in-sigma}),
we have $v_{m-2} \in \vars\bigl(s[x/t]\bigr)$.
Since, by Eq.~(\ref{eq: S-equiv-subst sigma'}),
$\bigl(y\mapsto s[x/t]\bigr) \in \sigma'$,
we have also $v_{m-2} \in \vars(y\sigma')$.
Moreover, by~(\ref{eq: S-equiv-subst vm-2-in-sigma}),
 $v_{m-2}\in \occ_{m-2}(\sigma,v)$ so that,
by the inductive hypothesis, we have that
$v_{m-2} \in \occ(\sigma', v)$.
Thus, by Eq.~(\ref{eq:property-of-occ}), as $v_m=y$, $v_m \in \occ(\sigma',v)$.

Conversely, we now prove that, for all $m\geq 0$,
\begin{equation}
\label{eq: S-equiv-subst occmsigma'-in-occbangsigma}
  \occ_m(\sigma', v) \sseq \occ(\sigma, v).
\end{equation}
The proof is again by induction on $m$.
As before, $\occ_0(\sigma', v) = \occ_0(\sigma, v) = \{v\}$
so that (\ref{eq: S-equiv-subst occmsigma'-in-occbangsigma}) holds for
$m=0$.
Suppose then that $m>0$ and
\(
 v_m \in \occ_m(\sigma', v).
\)
Then,
to prove  (\ref{eq: S-equiv-subst occmsigma'-in-occbangsigma}),
we must show that
\(
  v_m \in \occ(\sigma, v).
\)
By Definition~\ref{def:occurrence-functs},
there exists
\begin{equation}
\label{eq: S-equiv-subst vm-1-in-sigma'}
  v_{m-1}\in \vars (v_m\sigma') \inters \occ_{m-1}(\sigma', v).
\end{equation}
Hence, by the inductive hypothesis, $v_{m-1} \in \occ(\sigma, v)$.
If $v_{m-1} \in \vars(v_m\sigma)$ then,
by Eq.~(\ref{eq:property-of-occ}), we have $v_m \in \occ(\sigma, v)$.
Suppose now that $v_{m-1}\notin \vars (v_m\sigma)$.
Since, by~(\ref{eq: S-equiv-subst vm-1-in-sigma'}),
we have $v_{m-1} \in \vars(v_m\sigma')$,
it follows, using Eq.~(\ref{eq: S-equiv-subst sigma'}), that
$v_m = y$ and $v_{m-1} \in \vars(t) = \vars(x\sigma)$.
Hence, since
$v_{m-1} \in \occ(\sigma, v)$,
by Eq.~(\ref{eq:property-of-occ}), we have also $x \in \occ(\sigma, v)$.
Furthermore, $x\in \vars(y\sigma)$ so again,
by Eq.~(\ref{eq:property-of-occ}), as
$v_m=y$, $v_m \in \occ(\sigma, v)$.

Combining (\ref{eq: S-equiv-subst occmsigma-in-occbangsigma'}) and
(\ref{eq: S-equiv-subst occmsigma'-in-occbangsigma}) we obtain the result
that,
if $\sigma'$ is obtained from $\sigma$ by a single
\cS-step, then
$\occ(\sigma, v) = \occ(\sigma', v)$.
Thus, as $v \in \Vars$ was arbitrary,
$\alpha(\sigma, U) = \alpha(\sigma',U)$.

Suppose now that
$\sigma = \sigma_1 \Sstep \cdots \Sstep \sigma_n = \sigma'$.
If $n=1$, then $\sigma = \sigma'$.
If $n>1$, we have by the first part of the proof that,
for each $i=2$, \dots,~$n$,
 $\alpha(\sigma_{i-1},U) = \alpha(\sigma_i,U)$,
and hence the required result.
\end{proof}

\begin{proof}[Proof of Theorem~\ref{thm:equiv-ordered-vsubst}.]
By Theorem~\ref{thm:var-idem}, there exists $\sigma' \in \VSubst$ such that
$\sigma \Sstepstar \sigma'$ and, 
for any $\tau \sseq \sigma'$, $\tau \in \VSubst$.
Moreover,
by Theorem~\ref{thm:var-idem-equiv},
$\vars(\sigma) = \vars(\sigma')$ and
$T \entails \forall(\sigma \piff \sigma')$.
Thus, by Lemma~\ref{lem: S-equiv subst},
$\alpha(\sigma, U) = \alpha(\sigma',U)$.
\end{proof}

To prove Theorem~\ref{thm:equiv-subst}, we need to
show that the abstraction function $\alpha$
is invariant when we exchange equivalent variables 
to obtain an ordered substitution.
\begin{lem}
\label{lem: exchange-equiv-vars}
Suppose $\sigma \in \VSubst$, $v,w \in \Vars$ and
$(v \mapsto w) \in \sigma$.
Let $\rho = \{ v \mapsto w, w \mapsto v \}$ be a (circular) substitution
and define
$\sigma' = \rho \circ \sigma
         = \{\, x\rho \mapsto t\rho \mid x \mapsto t \in \sigma \,\}$.
Then
\begin{enumerate}
\item
$\sigma'\in \VSubst$,
\item
$\vars(\sigma) = \vars(\sigma')$,
\item
\label{sublem: exchange-equiv-vars: item4}
$\alpha(\sigma,U) = \alpha(\sigma',U)$, for all $U\in \wpf(\Vars)$, and
\item
\label{sublem: exchange-equiv-vars: item5}
$T\entails \forall (\sigma \piff \sigma')$, for any equality theory $T$.
\end{enumerate}
\end{lem}
 
\begin{proof}
Since $\sigma'$ is obtained from $\sigma$ by renaming variables
and $\sigma\in \VSubst$, we have also that $\sigma' \in \VSubst$.
In addition, $\vars(\sigma)\setdiff\{v,w\} = \vars(\sigma')\setdiff\{v,w\}$
so that,
since $(v \mapsto w) \in \sigma$
and $(w \mapsto v) \in \sigma'$,
we have $\vars(\sigma) = \vars(\sigma')$.

To prove property~\ref{sublem: exchange-equiv-vars: item4},
we have to show that, if
\[
  \alpha(\sigma,U) \defeq (\sh, U)
    \quad\text{ and }\quad
      \alpha(\sigma',U) \defeq (\sh', U),
\]
then $\sh = \sh'$.
By the hypothesis, for all $y \in \Vars$
we have $x \in \vars(y\sigma)$
if and only if $x\rho \in \vars(y\sigma')$.
As $\sigma, \sigma' \in \VSubst$,
we can use the alternative characterisation of $\occ$ given by
Lemma~\ref{lem:var-idem-occ-charac} and conclude that,
for each $x\in \Vars$,
$\occ(\sigma,x) = \occ(\sigma',x\rho)$.
Therefore $\sh \sseq \sh'$. 
The reverse inclusion follows by symmetry so that $\sh = \sh'$.

To prove property~\ref{sublem: exchange-equiv-vars: item5},
we first show by induction on the depth of $r \in \Terms$ that
\begin{equation}
\label{eq: exchange-equiv-vars: wv-equals-implies}
T \entails \forall \bigl((v=w) \pimplies (r=r\rho)\bigr).
\end{equation}
For the base case, $r$ has depth 1.
If $r$ is a constant or a variable other than $v$ or $w$, then
$r=r\rho$.
If $r=v$, then $r\rho =w$ and
 $T \entails \forall \bigl((v=w) \pimplies (v=w)\bigr)$.
Finally, if $r =w$, then $r\rho =v$ and we have, using the congruence
axioms,
that $T \entails \forall \bigl((v=w) \pimplies (w=v)\bigr)$.
For the inductive step, let $r = f(r_1,\ldots,r_n)$.
Then $r\rho = f(r_1\rho,\ldots,r_n\rho)$.
Thus, using the inductive hypothesis, for each $i=1$,~\ldots,~$n$,
$T \entails \forall \bigl((v=w) \pimplies (r_i=r_i\rho)\bigr)$.
Hence, by the congruence axioms,
(\ref{eq: exchange-equiv-vars: wv-equals-implies}) holds.

Note that $(v \mapsto w) \in \sigma$.
Thus, it follows from
(\ref{eq: exchange-equiv-vars: wv-equals-implies}) that,
for each $(x \mapsto t) \in \sigma$,
$T \entails \forall \bigl(\sigma \pimplies \{x=t, x=x\rho,
t=t\rho\}\bigr)$
and hence, using the congruence axioms,
$T \entails \forall \bigl(\sigma \pimplies \{x\rho=t\rho\}\bigr)$.
Thus,
$T \entails \forall (\sigma \pimplies \sigma')$.
Since $(w \mapsto v) \in \sigma'$,
the reverse implication follows by symmetry so that
$T \entails \forall (\sigma' \piff \sigma)$.
\end{proof}

\begin{lem}
\label{lem: order-videm}
Suppose $\sigma \in \VSubst$.
Then there exists $\sigma' \in \VSubst$ that is ordered such that 
$\vars(\sigma) = \vars(\sigma')$,
$\alpha(\sigma,U) = \alpha(\sigma',U)$, for all $U\in \wpf(\Vars)$, and
$T\entails \forall (\sigma \piff \sigma')$, for any equality theory $T$.
\end{lem}

\begin{proof}
The proof is by induction on the number $b \geq 0$
of the bindings $(v \mapsto w) \in \sigma$ such that
$w \in \param(\sigma)$ and
$w > v$ (the number of \emph{unordered} bindings).
For the base case, when $b = 0$, $\sigma$ is ordered and the 
result holds by taking $\sigma' = \sigma$.

For the inductive case, when $b>0$,
let $(v \mapsto w) \in \sigma$ be an unordered binding
and define $\rho = \{v \mapsto w, w\mapsto v\}$.
Then, by Lemma~\ref{lem: exchange-equiv-vars},
we have $\rho\circ\sigma \in \VSubst$, 
$\vars(\sigma) = \vars(\rho\circ\sigma)$,
$\alpha(\sigma,U) = \alpha(\rho\circ\sigma,U)$, 
for all $U\in \wpf(\Vars)$, and, finally,
$T\entails \forall (\sigma \piff \rho\circ\sigma)$, 
for any equality theory $T$.
In order to apply the inductive hypothesis to $\rho\circ\sigma$,
we must show that the number of unordered bindings
in $\rho\circ\sigma$ is less than $b$.
To this end, roughly speaking, we start showing that
any ordered binding in $\sigma$ is mapped by $\rho$ into
another ordered binding in $\rho\circ\sigma$, therefore proving
that the number of unordered bindings is not increasing.
There are three cases.
First, any ordered binding $(y \mapsto t) \in \sigma$
such that $t \notin \Vars$ is mapped by $\rho$ into
the binding $(y\rho \mapsto t\rho) \in (\rho\circ\sigma)$
which is clearly ordered, since $t\rho \notin \Vars$.
Second, consider any ordered binding $(y \mapsto z) \in \sigma$
such that $z \in \dom(\sigma)$.
Since $w \in \param(\sigma)$, we have $z \neq w$.
If also $z \neq v$
then we have $z\rho = z$ and $z \in \dom(\rho\circ\sigma)$;
otherwise $z=v$ so that $z\rho = w$ and,
as $(w \mapsto v) \in (\rho\circ\sigma)$,
$z\rho \in \dom(\rho\circ\sigma)$.
Thus, in either case, such a binding is mapped by $\rho$ into
the binding $(y\rho \mapsto z\rho) \in (\rho\circ\sigma)$
which is ordered since $z\rho \in \dom(\rho\circ\sigma)$.
Third, consider any ordered binding $(y \mapsto z) \in \sigma$
such that $z \in \param(\sigma)$ and $z < y$.
The ordering relation implies $y \neq v$ and
we also have $y \neq w$, since $w \in \param(\sigma)$.
Hence, we obtain $y\rho = y$.
Now, as $z \in \param(\sigma)$, $z \neq v$.
If $z \neq w$, then $z\rho = z$.
On the other hand, if $z = w$, then $z\rho = v$ so that $z\rho < z$.
Thus, in both cases, as $z < y$, $z\rho < y$.
and hence, $(y\rho \mapsto z\rho) \in (\rho\circ\sigma)$ is ordered.
Finally, to show that the number of unordered bindings is strictly
decreasing, we note that
the unordered binding $(v \mapsto w) \in \sigma$
is mapped by $\rho$ into
the binding $(w \mapsto v) \in (\rho\circ\sigma)$,
which is ordered.

Therefore, by applying the inductive hypothesis, there exists
a substitution $\sigma'$ such that
$\sigma' \in \VSubst$ is ordered,
$\vars(\rho\circ\sigma) = \vars(\sigma')$,
$\alpha(\rho\circ\sigma,U) = \alpha(\sigma',U)$, 
for all $U\in \wpf(\Vars)$, and
$T\entails \forall (\rho\circ\sigma \piff \sigma')$, 
for any equality theory $T$.
Then the required result follows by transitivity.
\end{proof}

\begin{proof}[Proof of Theorem~\ref{thm:equiv-subst}.]
By Theorem~\ref{thm:equiv-ordered-vsubst},
we can assume that $\sigma,\sigma' \in \VSubst$,
$T \entails \forall(\sigma \piff \sigma')$
and, for any $U \in \wpf(\Vars)$,
$\alpha(\sigma, U) = \alpha(\sigma', U)$.
By Lemma~\ref{lem: order-videm},
we can assume that $\sigma,\sigma'$
are also ordered substitutions so that, 
by Lemma~\ref{lem: var-idem-equiv-dom},
$\dom(\sigma') = \dom(\sigma)$.

To prove the result we need to show that,
for all $v\in \Vars$,
we have both
$\occ(\sigma, v) \sseq \occ(\sigma', v)$
and
$\occ(\sigma', v) \sseq \occ(\sigma, v)$.
We just prove the first of these as the other case is symmetric.

Suppose that $w \in \Vars$ and that
$v\in \vars(w\sigma)\setdiff\dom(\sigma)$.
Then, using the alternative characterisation of $\occ$
for variable-idempotent substitutions given by
Lemma~\ref{lem:var-idem-occ-charac},
we just have to show that
$v\in \vars(w\sigma')\setdiff\dom(\sigma')$.

By Lemma~\ref{lem: var-idem-equations}
(replacing $\tau$ by $\sigma$, $\sigma$ by $\sigma'$ and $s=t$ by $w=w$), 
we have that there exists
$z \in \vars(w\sigma')\setdiff \dom(\sigma')$ such that
$v \in \vars(z\sigma)$.
Thus as $\dom(\sigma') = \dom(\sigma)$,
$z \notin \dom(\sigma)$, and hence, $v = z$ so that
$v \in \vars(w\sigma')\setdiff\dom(\sigma')$, as required.
\end{proof}

\section{Abstract Unification}
\label{sec: abstract unify}

The operations of abstract unification
together with statements of the
main results are presented here in three stages.
In the first two stages, we consider substitutions containing
just a single binding.
For the first, it is assumed that the set of
variables of interest is fixed so that the
definition is based on the $\SH$ domain.
Then, in the second, using the $\SSl$ domain,
the definition is extended to allow
for the introduction of new variables in the binding.
The final stage extends this definition further to deal with arbitrary
substitutions.

\subsection{Abstract Operations for Sharing Sets}
\label{subsec: operations for sets}

The abstract unifier $\amgu$ abstracts the effect of a single binding on
an element of the $\SH$ domain.
For this we need some ancillary definitions.

\begin{defn} \summary{(Auxiliary functions.)}
\label{def:aux-funcs}
The \emph{closure under union} function
(also called \emph{star-union}),
$\fund{(\cdot)^\star}{\SH}{\SH}$,
is given,
for each $\sh \in \SH$, by
\[
  \sh^\star
    \defeq
      \bigl\{\,
        S \in \SG
      \bigm|
        \exists n \geq 1
          \st
            \exists S_1, \ldots, S_n \in \sh
              \st S = S_1 \union \cdots \union S_n
      \,\bigr\}.
\]
For each $\sh \in \SH$ and each $V \in \wpf(\Vars)$,
the extraction of the
\emph{relevant component of $\sh$ with respect to $V$}
is encoded by
$\fund{\rel}{\wpf(\Vars)\times\SH}{\SH}$
defined as
\[
  \rel(V, \sh)
    \defeq
      \{\, S \in \sh \mid S \inters V \neq \emptyset \,\}.
\]
For each $\sh_1, \sh_2 \in \SH$,
the \emph{binary union} function
$\fund{\bin}{\SH\times\SH}{\SH}$
is given by
\[
  \bin(\sh_1, \sh_2)
    \defeq
      \{\,
        S_1 \union S_2
      \mid
        S_1 \in \sh_1,
        S_2 \in \sh_2
      \,\}.
\]
\end{defn}

\begin{defn} \summary{($\amgu$.)}
\label{def:amgu}
The function $\fund{\amgu}{\SH\times\Bind}{\SH}$
captures the effects of a binding
on an $\SH$ element.
Suppose $x \in \Vars$, $r \in \Terms$, and $\sh \in \SH$.
Let
\begin{align*}
  A &\defeq \rel\bigl(\{x\}, \sh\bigr), \\
  B &\defeq \rel\bigl(\vars(r), \sh\bigr).\\
\intertext{%
  Then
}
  \amgu(\sh, x \mapsto r)
    &\defeq
      \bigl(\sh \setdiff (A \union B)\bigr)
        \union \bin(A^\star, B^\star).
\end{align*}
\end{defn}
The following soundness result for $\amgu$
is proved in Section~\ref{sec:amgu results}.
\begin{thm}
\label{thm: soundness of amgu}
Let $T$ be a syntactic equality theory,
$(\sh,U) \in \SSl$ an abstract description
and $\{x \mapsto r\}, \sigma \in \RSubst$
 such that $\vars(x\mapsto r) \union \vars(\sigma) \sseq U$.
Suppose that there exists a most general solution $\mu$ for
$\bigl(\{x = r\} \union \sigma\bigr)$ in $T$.
Then
\[
  \alpha(\sigma, U) \leqSSl (\sh, U)
    \implies
      \alpha(\mu, U)
        \leqSSl \bigl(\amgu(\sh, x \mapsto r), U\bigr).
\]
\end{thm}

The following theorems,
proved  in Section~\ref{sec:amgu results},
show that $\amgu$ is idempotent and commutative.
\begin{thm}
\label{thm: idempotence of amgu}
Let $\sh \in \SH$ and $(x \mapsto r) \in \Bind$.
Then
\[
  \amgu(\sh, x \mapsto r)
    = \amgu\bigl(\amgu(\sh, x \mapsto r), x \mapsto r\bigr).
\]
\end{thm}

\begin{thm}
\label{thm: commutativity of amgu}
Let $\sh \in \SH$ and $(x \mapsto r), (y \mapsto t) \in \Bind$.
Then
\[
    \amgu\bigl(\amgu(\sh, x \mapsto r), y \mapsto t\bigr) =
        \amgu\bigl(\amgu(\sh, y \mapsto t), x \mapsto r\bigr).
\]
\end{thm}

\subsection{Abstract Operations for Sharing Domains}
\label{subsec: operations for domains}

The definitions and results of Section~\ref{subsec: operations for sets}
can be lifted to apply to the proper set-sharing domain.

\begin{defn} \summary{($\Amgu$.)}
\label{def:Amgu}
The operation $\fund{\Amgu}{\SSl\times\Bind}{\SSl}$ extends the
$\SSl$ description it takes as an argument
to the set of variables occurring in the binding it is given
as the second argument. Then it applies $\amgu$.
Formally:
\begin{align*}
  U'
    &\defeq
      \vars(x \mapsto r) \setdiff U, \\
  \Amgu\bigl((\sh, U), x \mapsto r\bigr)
    &\defeq
      \biggl(
        \amgu\Bigl(
               \sh \union
                 \bigl\{\,
                   \{u\}
                 \bigm|
                   u \in U'
                 \,\bigr\},
               x\mapsto r
             \Bigr),
        U \union U'
      \biggr).
\end{align*}
\end{defn}

The results for $\amgu$ can easily be extended to apply to $\Amgu$
giving us the following corollaries.
\pagebreak[3]
\begin{cor}
\label{cor: soundness of Amgu}
Let $T$ be a syntactic equality theory,
$(\sh,U) \in \SSl$ and
$\{x \mapsto r\}, \sigma \in \RSubst$
such that $\vars(\sigma) \sseq U$.
Suppose  there exists a most general solution $\mu$ for
$\bigl(\{x = r\} \union \sigma\bigr)$ in $T$.
Then
\[
  \alpha(\sigma, U) \leqSSl (\sh,U)
    \implies
      \alpha\bigl(
              \mu, U \union \vars(x\mapsto r)
            \bigr)
        \leqSSl \Amgu\bigl((\sh,U), x \mapsto r\bigr).
\]
\end{cor}

\begin{cor}
\label{cor: idempotence of Amgu}
Let $\sh \in \SH$ and $(x \mapsto r) \in \Bind$.
Then
\[
  \Amgu\bigl((\sh,U), x \mapsto r\bigr)
    = \Amgu\Bigl(\Amgu\bigl((\sh,U), x \mapsto r\bigr), x \mapsto r\Bigr).
\]
\end{cor}
\begin{cor}
\label{cor: commutativity of Amgu}
Let $\sh \in \SH$ and $(x \mapsto r), (y \mapsto t) \in \Bind$.
Then
\begin{multline*}
  \Amgu\Bigl(\Amgu\bigl((\sh,U), x \mapsto r\bigr), y \mapsto t\Bigr) \\
    = \Amgu\Bigl(\Amgu\bigl((\sh,U), y \mapsto t\bigr), x \mapsto r\Bigr).
\end{multline*}
\end{cor}

\subsection{Abstract Unifiers for Sharing}
\label{subsec: unifiers for sharing}

We now extend the above definitions and results for a single binding to any
substitution.

\begin{defn} \summary{($\aunify$.)}
\label{def:aunify}
The function $\fund{\aunify}{\SSl\times\RSubst}{\SSl}$
generalizes $\Amgu$ to any substitution $\mu\in \RSubst$ in the context of some
syntactic equality theory $T$:
If we have $(\sh, U) \in \SSl$,
then
\begin{align*}
  \aunify\bigl((\sh, U), \emptyset\bigr)
    &\defeq
      (\sh, U); \\
\intertext{%
if  $\mu$ is satisfiable in $T$ and $(x\mapsto r) \in \mu$,
}
  \aunify\bigl((\sh, U), \mu \bigr)
    &\defeq
      \aunify\Bigl(\bigl(\Amgu(\sh, U), x \mapsto r\bigr),
                 \mu \setdiff \{x \mapsto r\} \Bigr); \\
\intertext{%
and, if  $\mu$ is not satisfiable in $T$,
}
  \aunify\bigl((\sh, U), \mu\bigr)
    &\defeq
      \bot.\\
\intertext{%
For the distinguished elements $\bot$ and $\top$ of $\SSl$,
}
  \aunify(\bot, \mu)
    &\defeq
      \bot, \\
  \aunify(\top, \mu)
    &\defeq
      \top.
\end{align*}
\end{defn}

As a result of Corollary~\ref{cor: commutativity of Amgu},
$\Amgu$ and $\aunify$ commute.

\begin{lem}
\label{lem: commutativity of aunify/Amgu}
Let $(\sh,U) \in \SSl$, $\nu\in \RSubst$ and $(y \mapsto t) \in \Bind$.
Then
\[
  \aunify\Bigl(\Amgu\bigl((\sh, U), y \mapsto t\bigr), \nu\Bigr)
    =
      \Amgu\Bigl(\aunify\bigl((\sh, U), \nu\bigr), y \mapsto t\Bigr).
\]
\end{lem}
As a consequence of this and Corollaries~\ref{cor: soundness of Amgu},
\ref{cor: idempotence of Amgu} and~\ref{cor: commutativity of Amgu},
we have the following soundness, idempotence
and commutativity results required for $\aunify$ to be sound and
well-defined.

\begin{thm}
\label{thm: soundness of aunify}
Let $T$ be a syntactic equality theory,
$(\sh,U) \in \SSl$ and
$\sigma,\nu \in \RSubst$
such that $\vars(\sigma) \sseq U$.
Suppose also that there exists a most general solution $\mu$ for
$(\nu \union \sigma)$ in $T$.
Then
\[
  \alpha(\sigma, U) \leqSSl (\sh,U)
    \implies
      \alpha\bigl(
              \mu, U \union \vars(\nu)
            \bigr)
        \leqSSl \aunify\bigl((\sh,U), \mu\bigr).
\]
\end{thm}
This theorem shows also
that it is safe for the analyzer to perform part or all
of the concrete unification algorithm before computing $\aunify$.

\begin{thm}
\label{thm: idempotence of aunify}
Let $(\sh,U) \in \SSl$ and $\nu \in \RSubst$.
Then
\[
  \aunify\bigl((\sh,U), \nu\bigr) =
  \aunify\Bigl(\aunify\bigl((\sh,U),\nu\bigr), \nu\Bigr).
\]
\end{thm}

\begin{thm}
\label{thm: commutativity of aunify}
Let $(\sh,U) \in \SSl$ and $\nu_1, \nu_2 \in \RSubst$.
Then
\[
  \aunify\Bigl(\aunify\bigl((\sh,U), \nu_1\bigr), \nu_2\Bigr)
    =
  \aunify\Bigl(\aunify\bigl((\sh,U), \nu_2\bigr), \nu_1\Bigr).
\]
\end{thm}
The proofs of all these results are
in Section~\ref{sec:aunify results}.

\subsection{Proofs of Results for Sharing-Sets}
\label{sec:amgu results}

In the proofs we use the fact that $(\cdot)^\star$ and $\rel$ are
monotonic
so that
\begin{align}
\label{eq:emi-star}
  \sh_1 \sseq \sh_2
    &\implies
      \sh_1^\star \sseq \sh_2^\star,\\
\label{eq:emi-rel}
  \sh_1 \sseq \sh_2
    &\implies
      \rel(\sh_1,U) \sseq \rel(\sh_2,U).
\end{align}
We will also use the fact that  $(\cdot)^\star$ is idempotent.

Let $t_1$, \dots,~$t_n$ be terms.
For the sake of brevity we will use the notation
$v_{t_1 \cdots t_n}$ to denote $\bigunion_{i=1}^n \vars(t_i)$.
In particular, if $x$ and $y$ are variables,
and $r$ and $t$ are terms, we will use the following definitions:
\begin{align*}
  v_x &\defeq \{x\},
    & v_y &\defeq \{y\}, \\
  v_r &\defeq \vars(r),
    & v_t &\defeq \vars(t), \\
  v_{xr} &\defeq v_x \union v_r,
    & v_{yt} &\defeq v_y \union v_t.
\end{align*}

\begin{defn} \summary{($\mathord{\irel}$.)}
Suppose $V \in \wpf(\Vars)$ and $\sh \in \SH$. Then
\[
    \irel(V, \sh) \defeq \sh \setdiff \rel(V, \sh).
\]
\end{defn}

Notice that
if $S \in \irel(V, \sh)$ then $S \inters V = \emptyset$.
Conversely,
if $S \in \sh$ and $S \inters V = \emptyset$ then $S \in \irel(V, \sh)$.
The following definition of $\amgu$
is clearly equivalent to the one given in Definition~\ref{def:amgu}:
for each variable $x$, each term $r$,
and each $\sh \in \SH$,
\begin{equation}
\label{eq:amgu-in-terms-of-irel}
  \amgu(\sh, x \mapsto r)
    \,\defeq\,
      \irel(v_{xr}, \sh)
        \union \bin\bigl(\rel(v_x, \sh)^\star, \rel(v_r, \sh)^\star\bigr).
\end{equation}

\begin{proof*}[Proof of Theorem~\ref{thm: soundness of amgu}.]
We first prove the result under the assumption that 
$\alpha(\sigma, U) = (\sh,U)$.
We do this in two parts. 
In the first, we partition $\sigma$ into two substitutions one of 
which, called $\sigma^-$, is the same as $\sigma$ when
$\sigma$ and $\mu$ are idempotent.
We construct a new substitution $\nu$ which, 
in the case that $\sigma$ and $\mu$ are idempotent, is a 
most general solution for $x\sigma = r\sigma$. 
Finally we compose $\nu$ with $\sigma^-$ to 
define a substitution that has the same abstraction as $\mu$ 
but with a number of useful properties including that of variable-idempotence.
In the second part, we use this composed substitution in place of $\mu$ to 
prove the result.

\textbf{Part 1.}
By Theorem~\ref{thm:equiv-ordered-vsubst},
we can assume that 
\begin{equation}
\label{eq:soundness-proof-sigma-videm}
\sigma \in\VSubst 
\end{equation}
 and that all subsets of $\sigma$ are in $\VSubst$.
Let $\sigma^\circ,\sigma^- \in \RSubst$
be defined such that
\begin{align}
\label{eq:soundness-proof-sigma--sigmacirc-def}
 \sigma^- \union \sigma^\circ &= \sigma,\\
\label{eq:soundness-proof-sigmacirc-def}
 \dom(\sigma^\circ) &= \dom(\sigma) \inters 
     \bigcup_{i\geq 1}\vars(x\sigma^i = r\sigma^i),\\
\label{eq:soundness-proof-sigma--def}
 \dom(\sigma^-)\inters \dom(\sigma^\circ) &= \emptyset.
\end{align} 
Then, it follows from the above assumption on subsets of $\sigma$ that
\begin{equation}
\label{eq:soundness-proof-allsigma-videm}
\sigma^-\in \VSubst, \quad \sigma^\circ \in \VSubst.
\end{equation}
Now, suppose $z \in \vars(\sigma^\circ)\setdiff\dom(\sigma^\circ)$.
Then $z\in \vars(y\sigma^\circ)$ for some $y\in \dom(\sigma^\circ)$.
Thus, by~(\ref{eq:soundness-proof-sigmacirc-def}), 
 for some $j \geq 2$, 
$z \in \vars(x\sigma^j = r\sigma^j)\setdiff\dom(\sigma^\circ)$
and, again
by~(\ref{eq:soundness-proof-sigmacirc-def}),
 $z\notin\dom(\sigma)$
so that, by~(\ref{eq:soundness-proof-sigma-videm}),
$z \in \vars(x\sigma = r\sigma)$.
Therefore, as $z$ was an arbitrary variable in 
$\vars(\sigma^\circ)\setdiff\dom(\sigma^\circ)$,
\begin{gather}
\label{eq:soundness-proof-vars-sigmacirc-sseq-dom}
\vars(\sigma^\circ) \sseq 
   \bigl(\vars(x\sigma = r\sigma) \union \dom(\sigma^\circ)\bigr).\\
\intertext{%
It follows from~(\ref{eq:soundness-proof-sigmacirc-def})
that $\dom(\sigma) \inters 
     \vars(x\sigma = r\sigma) \sseq \dom(\sigma^\circ)$ so that,
by~(\ref{eq:soundness-proof-sigma--def}) 
}
\label{eq:soundness-proof-dom-sigma--xrsigma}
\dom(\sigma^-) \inters \vars(x\sigma = r\sigma) = \emptyset.\\
\intertext{%
Hence, by~(\ref{eq:soundness-proof-sigma--def}) 
and~(\ref{eq:soundness-proof-vars-sigmacirc-sseq-dom}), we have
}
\label{eq:soundness-proof-dom-sigma--sigmacirc}
\dom(\sigma^-) \inters \vars(\sigma^\circ) = \emptyset.
\end{gather}

Let $\nu\in \RSubst$ be a most general solution for
$\{x\sigma = r\sigma\}\union \sigma^\circ$ in $T$
so that
\begin{align}
\label{eq:soundness-proof-nu-equiv}
  T &\entails 
     \forall\bigl(\nu \piff \{x\sigma = r\sigma\}\union \sigma^\circ\bigr),\\
\label{eq:soundness-proof-nu-relevant}
\vars(\nu) &\sseq \bigl(\vars(x\sigma = r\sigma) \union \vars(\sigma^\circ)\bigr).
\end{align}
By Theorem~\ref{thm:equiv-ordered-vsubst},
we can assume that 
\begin{equation}
\label{eq:soundness-proof-nu-videm}
\nu \in \VSubst.
\end{equation}
By~(\ref{eq:soundness-proof-dom-sigma--xrsigma}),
(\ref{eq:soundness-proof-dom-sigma--sigmacirc}), and 
(\ref{eq:soundness-proof-nu-relevant}), we have
\begin{equation}
\label{eq:soundness-proof-dom-sigma-}
\dom(\sigma^-) \inters \vars(\nu) = \emptyset.
\end{equation}
Therefore, as $\sigma^-,\nu \in \VSubst$ 
(by~(\ref{eq:soundness-proof-allsigma-videm}) and 
(\ref{eq:soundness-proof-nu-videm})),
we can use Lemma~\ref{lem: properties-of-composition}
to obtain
the following properties for~$\nu\circ\sigma^-$.
\begin{gather}
\label{eq:soundness-proof-nu-sigma-equiv}
  T \entails  
    \forall\bigl((\nu\circ\sigma^-) \piff (\nu \union \sigma^-)\bigr),\\
\label{eq:soundness-proof-nu-sigma-doms}
  \dom(\nu\circ\sigma^-) = \dom(\nu \union \sigma^-),\\
\label{eq:soundness-proof-nu-sigma-videm}
\nu\circ\sigma^- \in \VSubst.
\end{gather}

Now we have
\begin{align}
\notag
T &\entails \forall\bigl(\mu \piff \{x = r\} \union \sigma\bigr) \\
\notag
    &\just{by hypothesis}\\
\notag
T &\entails \forall\bigl(\mu \piff
  \{x\sigma = r\sigma\} \union \sigma\bigr)  \\
\notag
    &\just{by Lemma~\ref{lem: applic} and the congruence axioms}\\
\notag
T &\entails \forall\bigl(\mu \piff \nu \union \sigma^-\bigr) \\
\notag
    &\just{by~(\ref{eq:soundness-proof-sigma--sigmacirc-def}) and
              (\ref{eq:soundness-proof-nu-equiv})}\\
\label{eq:soundness-proof-mu-nu-sigma--equiv}
T &\entails \forall\bigl(\mu \piff \nu \circ \sigma^-\bigr) \\
\notag
    &\just{by~(\ref{eq:soundness-proof-nu-sigma-equiv})}.
\end{align}
Therefore, by Theorem~\ref{thm:equiv-subst}, 
\begin{equation}
\label{eq:soundness-proof-alpha-mu'}
\alpha(\mu, U) = \alpha(\nu\circ\sigma^-, U).
\end{equation}

\textbf{Part 2.}
To prove the result under the assumption that
$\alpha(\sigma, U) = (\sh,U)$, we define $\sh'\in \SH$ so that
\begin{equation}
\label{eq:soundness-proof-define-sh'}
\alpha(\mu, U) = (\sh',U).
\end{equation}
Then, by~(\ref{eq:soundness-proof-alpha-mu'}),
$\alpha(\nu\circ\sigma^-, U) = (\sh',U)$.
We show that $\sh' \sseq \amgu(\sh, x\mapsto r)$.
If $\sh' = \emptyset$, there is nothing to prove.
Therefore, we assume that there exists $S \in \sh'$ so that
$S\neq \emptyset$
and, for some $v\in \Vars$,
\begin{align}
\label{eq:soundness-proof-v-notin-dom-mu}
  v &\notin \dom(\nu\circ\sigma^-),\\
\label{eq:soundness-proof-1st-def-of-S}
  S &\defeq \occ(\nu\circ\sigma^-,v).
\end{align}
Note that~(\ref{eq:soundness-proof-nu-sigma-doms}) and
(\ref{eq:soundness-proof-v-notin-dom-mu}) imply that
\begin{equation}
\label{eq:soundness-proof-nu-dom}
v \notin \dom(\nu), \qquad  v \notin \dom(\sigma^-).
\end{equation}

Let
\begin{align}
\label{eq:soundness-proof-2nd-def-of-S}
   S' &\defeq \bigunion \bigl\{\, \occ(\sigma, y) \bigm|
            y \in \occ(\nu,v) \,
                 \bigr\}.
\end{align}
We show that
\begin{equation}
\label{eq:soundness-proof-S=S'}
S = S'.
\end{equation}
By (\ref{eq:soundness-proof-sigma-videm}),
 (\ref{eq:soundness-proof-nu-videm}) and
(\ref{eq:soundness-proof-nu-sigma-videm}), 
$\sigma,\nu,\nu\circ\sigma^- \in \VSubst$ and, 
by~(\ref{eq:soundness-proof-v-notin-dom-mu})
and~(\ref{eq:soundness-proof-nu-dom}),
$v \notin \dom(\nu\circ\sigma^-)$ and $v \notin \dom(\nu)$.
Thus, it follows from Lemma~\ref{lem:var-idem-occ-charac} with 
(\ref{eq:soundness-proof-1st-def-of-S}) and
(\ref{eq:soundness-proof-2nd-def-of-S}),
that it suffices to show that,
for each $w \in \Vars$,
$v\in \vars(w\sigma^-\nu)$
if and only if
there exists
$z \in \vars(w\sigma) \setminus \dom(\sigma)$
such that
$v\in \vars(z\nu)$. 

First,
we suppose that $v\in\vars(w\sigma^-\nu)$. Thus,
there exists $y \in \vars(w\sigma^-)$ such that $v \in \vars(y\nu)$.
Since $\sigma^\circ,\nu \in \VSubst$ (by~(\ref{eq:soundness-proof-allsigma-videm}) and 
(\ref{eq:soundness-proof-nu-videm})),
$T\entails \forall(\nu \pimplies \sigma^\circ)$ 
(by~(\ref{eq:soundness-proof-nu-equiv})), 
$v\notin \dom(\nu)$
(by~(\ref{eq:soundness-proof-nu-dom}))
 and
$T\entails \forall\bigl(\nu \pimplies (y\nu = y)\bigr)$ 
(using Lemma~\ref{lem: applic}), we can apply 
Lemma~\ref{lem: var-idem-equations}
(replacing $\tau$ by $\nu$, $\sigma$ by $\sigma^\circ$ and $s=t$ by $y\nu=y$)
so that there exists 
$z \in \vars(y\sigma^\circ)\setdiff\dom(\sigma^\circ)$ such that 
$v \in \vars(z\nu)$.
We want to show that
$z \in \vars(w\sigma)\setdiff\dom(\sigma)$.
Now either $z\in \dom(\nu)$ or $z=v$ so that,
by
(\ref{eq:soundness-proof-dom-sigma-}) (if $z\in \dom(\nu)$) or 
(\ref{eq:soundness-proof-nu-dom}) (if $z=v$), 
$z\notin \dom(\sigma^-)$.
However, $z\notin \dom(\sigma^\circ)$, 
so that, by~(\ref{eq:soundness-proof-sigma--sigmacirc-def}),
$z\notin \dom(\sigma)$.
Thus, it remains to prove that $z \in \vars(w\sigma)$.
Now, as $y\in\vars(w\sigma^-)$ and
$z \in \vars(y\sigma^\circ)$, 
we have $z \in \vars(w\sigma^-\sigma^\circ)$.
So we must show that
$\vars(w\sigma^-\sigma^\circ)\setdiff\dom(\sigma) \sseq 
    \vars(w\sigma)$.
To see this note that, if $w \notin \dom(\sigma^-)$, then
$w\sigma^- = w$ and, by~(\ref{eq:soundness-proof-sigma--sigmacirc-def}), 
$w\sigma^\circ = w\sigma$ so that
$w\sigma^-\sigma^\circ = w\sigma$. 
On the other hand, if $w \in \dom(\sigma^-)$, then,
by~(\ref{eq:soundness-proof-sigma--sigmacirc-def}),
$w\sigma^- = w\sigma$ so that 
$w\sigma^-\sigma^\circ = w\sigma\sigma^\circ$
Now, as 
$\sigma \in \VSubst$ and $\sigma^\circ \sseq \sigma$ 
(by~(\ref{eq:soundness-proof-sigma-videm}) and
(\ref{eq:soundness-proof-sigma--sigmacirc-def})),
we can apply Lemma~\ref{lem:var-idem-subset-property} so that
$\vars(w\sigma\sigma^\circ)\setdiff\dom(\sigma) \sseq  \vars(w\sigma)$.
Hence,
$\vars(w\sigma^-\sigma^\circ)\setdiff\dom(\sigma) \sseq 
    \vars(w\sigma)$.

Secondly, suppose there exists
$z \in \vars(w\sigma) \setminus \dom(\sigma)$
such that $v\in \vars(z\nu)$.
Then $v\in \vars(w\sigma\nu)$.
We need to show that $v\in \vars(w\sigma^-\nu)$.
By Eq.~(\ref{eq:soundness-proof-sigma--sigmacirc-def}),
if $w\in\dom(\sigma^-)$, 
then $w\sigma^-\nu = w\sigma\nu$ so that
$v \in \vars(w\sigma^-\nu)$.
On the other hand,
if $w\notin\dom(\sigma^-)$, then
again, by~(\ref{eq:soundness-proof-sigma--sigmacirc-def}),
$v \in \vars(w\sigma^\circ\nu)$.
Moreover, $w = w\sigma^-$ so that,
by~(\ref{eq:soundness-proof-nu-equiv}) and
 Lemma~\ref{lem: applic} with the congruence axioms, 
 $T \entails \forall\bigl(\nu \pimplies (w\sigma^\circ\nu = w\sigma^-)\bigr)$.
Hence, since $\nu \in \VSubst$ (by~(\ref{eq:soundness-proof-nu-videm}))
 and $v\notin \dom(\nu)$ (by~(\ref{eq:soundness-proof-nu-dom})), 
we can apply Lemma~\ref{lem: var-idem-equations}
(replacing $\tau$ by $\nu$, $\sigma$ by the empty substitution and 
$s=t$ by $w\sigma^\circ\nu = w\sigma^-$)
 and obtain
$v \in \vars(w\sigma^-\nu)$.

Therefore, as a consequence of the previous two paragraphs,
for each $w \in \Vars$,
we have $v\in \vars(w\sigma^-\nu)$
if and only if
there exists
$z \in \vars(w\sigma) \setminus \dom(\sigma)$
such that
$v\in \vars(z\nu)$.
It therefore follows that 
Eq.~(\ref{eq:soundness-proof-S=S'}) holds.

Let
\begin{align}
\label{eq:soundness-proof-S_x}
  S_x
    &\defeq
      \bigunion \Bigl(\bigl\{\, \occ(\sigma, y)
                 \bigm| y \in \occ(\nu, v)
                \,\bigr\} \inters \rel(v_x,\sh)\Bigr), \\
\label{eq:soundness-proof-S_r}
  S_r
    &\defeq
      \bigunion \Bigl(\bigl\{\, \occ(\sigma, y)
                \bigm| y \in \occ(\nu, v)
                \,\bigr\} \inters \rel(v_r,\sh)\Bigr), \\
\label{eq:soundness-proof-S_0}
  S_0 &\defeq
      \bigunion \Bigl(\bigl\{\, \occ(\sigma, y)
                \bigm| y \in \occ(\nu, v)
                \,\bigr\} \inters \irel(v_{xr},\sh)\Bigr).
\end{align}
Note that by~(\ref{eq:soundness-proof-2nd-def-of-S}),~(\ref{eq:soundness-proof-S=S'}) and
the fact that
\[
  \irel(v_{xr},\sh)
     =
       \sh \setdiff \bigl(\rel(v_x,\sh) \union \rel(v_r,\sh)\bigr),
\]
we have
\begin{equation}
\label{eq:soundness-proof-S-without-xr}
S_0 = S \setdiff (S_x \union S_r).
\end{equation}
We now consider the two cases  $S_0 \neq \emptyset$ and $S_0 = \emptyset$
separately.

Consider first the case when $S_0 \neq \emptyset$.
Then, by~(\ref{eq:soundness-proof-S_0}),
 for some $y\in \Vars$,
\begin{align}
\label{eq:soundness-proof-occ-y-v}
y &\in \occ(\nu,v),\\
\label{eq:soundness-proof-occ-y-irel}
\occ(\sigma, y) &\in \irel(v_{xr},\sh).
\end{align}
Thus, by Lemma~\ref{lem:var-idem-dom-charac},
$y \notin \dom(\sigma)$ and hence, 
by~(\ref{eq:soundness-proof-sigma--sigmacirc-def}), 
$y \notin \dom(\sigma^\circ)$.
Also, by~(\ref{eq:soundness-proof-occ-y-irel}), 
$\occ(\sigma,y) \inters  v_{xr} = \emptyset$.
Thus as $\sigma \in \VSubst$ (by~(\ref{eq:soundness-proof-sigma-videm})) 
we can use Lemma~\ref{lem:var-idem-occ-charac} to see that, 
for each $w\in v_{xr}$, $y \notin \vars(w\sigma)$
and hence,
$y \notin \vars(x\sigma = r\sigma)$.
Therefore, by
(\ref{eq:soundness-proof-vars-sigmacirc-sseq-dom}) 
and~(\ref{eq:soundness-proof-nu-relevant}),
$y \notin \vars(\nu)$.
As $\nu \in \VSubst$ 
(by~(\ref{eq:soundness-proof-nu-videm})), we can apply
 Lemma~\ref{lem:var-idem-occ-charac} to both $\occ(\nu,y)$ and $\occ(\nu,v)$.
Thus, as $y \notin \vars(\nu)$,
$\occ(\nu,y) = \{y\}$
and also (using~(\ref{eq:soundness-proof-occ-y-v})) $v=y$
so that $\occ(\nu, v) = \{v\}$.
It therefore follows from (\ref{eq:soundness-proof-2nd-def-of-S}) and
(\ref{eq:soundness-proof-S=S'}) that
$S = \occ(\sigma,v)$ and hence
from~(\ref{eq:soundness-proof-occ-y-irel}), that
\begin{equation}
\label{eq:soundness-proof-S_0-nonempty}
S \in \irel(v_{xr},\sh).
\end{equation}

Now consider the case when $S_0 = \emptyset$.
By~(\ref{eq:soundness-proof-S-without-xr}),
and the assumption that $S\neq\emptyset$,
\begin{align}
\label{eq:soundness-proof-S_0=emptyset}
  S &= S_x \union S_r \neq \emptyset.\\
\intertext{%
As a consequence of (\ref{eq:soundness-proof-S_x}) and
(\ref{eq:soundness-proof-S_r}),
}
\label{eq:soundness-proof-S_x-again}
  S_x &\in \rel(v_x, \sh)^\star \union \emptyset,\\
\label{eq:soundness-proof-S_r-again}
  S_r &\in \rel(v_r, \sh)^\star \union \emptyset. \\
\intertext{%
Now, by~(\ref{eq:soundness-proof-S_0=emptyset})
either $S_x\neq \emptyset$ or $S_r\neq \emptyset$.
We will show that both $S_x\neq \emptyset$ and $S_r\neq \emptyset$.
Suppose first that $S_x \neq \emptyset$.
Then, by~(\ref{eq:soundness-proof-S_x-again}), $x\in S_x$. 
Hence, by~(\ref{eq:soundness-proof-S_0=emptyset}), $x\in S$.
By~(\ref{eq:soundness-proof-1st-def-of-S}),
$x \in \occ(\nu\circ\sigma^-,v)$.
However, $\nu\circ\sigma^- \in \VSubst$ 
(by~(\ref{eq:soundness-proof-nu-sigma-videm}))
 so that we can apply Lemma~\ref{lem:var-idem-occ-charac}
to $\occ(\nu\circ\sigma^-,v)$ and obtain that 
$v \in \vars(x\sigma^-\nu)$.
By the definition of $\mu$ in the hypothesis 
and~(\ref{eq:soundness-proof-mu-nu-sigma--equiv}),
$T \entails \forall\bigl(\nu\circ\sigma^- \pimplies (x = r)\bigr)$ 
and hence, by Lemma~\ref{lem: applic} with the congruence axioms,
$T \entails \forall\bigl(\nu\circ\sigma^- \pimplies (x\sigma^-\nu = r)\bigr)$.
Thus, as $\nu\circ\sigma^- \in \VSubst$
(by~(\ref{eq:soundness-proof-nu-sigma-videm}))
and $v \notin \dom(\nu\circ\sigma^-)$
(by~(\ref{eq:soundness-proof-v-notin-dom-mu})),
we have, by Lemma~\ref{lem: var-idem-equations}
(replacing $\tau$ by $\nu\circ\sigma^-$, 
      $\sigma$ by the empty substitution and $s=t$ by $x\sigma^-\nu=r$),
 $v \in \vars(r\sigma^-\nu)$.
By re-applying Lemma~\ref{lem:var-idem-occ-charac}
to $\occ(\nu\circ\sigma^-,v)$,
it can be seen that, as
$v\notin \dom(\nu)$ (by~(\ref{eq:soundness-proof-v-notin-dom-mu})),
$v_r \inters \occ(\nu\circ\sigma^-,v) \neq \emptyset$.
Hence, by~(\ref{eq:soundness-proof-1st-def-of-S}),
$S \inters v_r \neq \emptyset$.
Thus, by~(\ref{eq:soundness-proof-2nd-def-of-S}) 
and~(\ref{eq:soundness-proof-S=S'}),
there exists a $y\in \occ(\nu,v)$ such that
$\occ(\sigma,y)\inters v_r \neq \emptyset$.
Therefore, by~(\ref{eq:soundness-proof-S_r}),
$S_r \inters v_r \neq \emptyset$
and so $S_r \neq \emptyset$.
Secondly, by a similar argument, if $S_r \neq \emptyset$ then we have
$S_x \neq \emptyset$.
Hence
$S_x \neq \emptyset$ and $S_r \neq \emptyset$.
So that, by~(\ref{eq:soundness-proof-S_x-again}) 
and~(\ref{eq:soundness-proof-S_r-again}),
  $S_x \in \rel(v_x, \sh)^\star$ and
  $S_r \in \rel(v_r, \sh)^\star$.
Therefore, we have, by~(\ref{eq:soundness-proof-S_0=emptyset}),
}
\label{eq:soundness-proof-S_x-and-S_r-nonempty}
    S
      &\in \bin\bigl(\rel(v_x, \sh)^\star,
                    \rel(v_r, \sh)^\star
              \bigr). \\
\intertext{%
Combining (\ref{eq:soundness-proof-S_0-nonempty}) when $S_0 \neq \emptyset$
and (\ref{eq:soundness-proof-S_x-and-S_r-nonempty}) when $S_0 = \emptyset$
we obtain
}
\notag
   S  &\in \irel(v_{xr},\sh) \union
            \bin\bigl(\rel(v_x, \sh)^\star,
                    \rel(v_r, \sh)^\star
                \bigr)\\
\intertext{%
and therefore, by~(\ref{eq:amgu-in-terms-of-irel}),
}
\notag
   S  &\in \amgu(\sh, x \mapsto r).
\end{align}
As a consequence, since $S$ was any set in $\sh'$,
we have $\sh' \sseq \amgu(\sh, x\mapsto r)$ and hence,
by~(\ref{eq:soundness-proof-define-sh'}),
\begin{equation}
\label{eq:soundness-proof-=-case}
\alpha(\mu, U) \leqSSl \bigl(\amgu(\sh, x \mapsto r), U\bigr).
\end{equation}

We now drop the assumption that $\alpha(\sigma, U) = (\sh,U)$ and
just assume the hypothesis of the theorem that
$\alpha(\sigma,U) \leqSSl (\sh,U)$.
Suppose $\alpha(\sigma,U) = (\sh_1,U)$. Then $\sh_1 \sseq \sh$.
It follows from Definition~\ref{def:amgu} that
$\amgu$ is monotonic on its first argument so that
\[
   \amgu(\sh_1, x \mapsto r) \sseq \amgu(\sh, x \mapsto r).
\]
Thus, by~(\ref{eq:soundness-proof-=-case}) (replacing $\sh$ by $\sh_1$),
we obtain the required result
\[
  \alpha(\mu, U) \leqSSl \bigl(\amgu(\sh, x \mapsto r), U\bigr).
\mathproofbox
\]
\end{proof*}

\begin{lem}
\label{lem:bin-star}
For each $\sh_1,\sh_2 \in \SH$, we have
\[
  \bin(\sh_1, \sh_2)^\star = \bin(\sh_1^\star, \sh_2^\star).
\]
\end{lem}
\begin{proof}
Suppose $S \in \SG$.
Then $S \in \bin(\sh_1, \sh_2)^\star$ means that, for some $n \in \Nset$,
there exist sets $R_1, \ldots, R_n \in \sh_1$
and $T_1, \ldots, T_n \in \sh_2$ such that
$S = (R_1 \union T_1) \union\cdots\union (R_n \union T_n)$.
Thus $S = (R_1 \union\cdots\union R_n) \union (T_1 \union\cdots\union T_n)$.
However $R_1 \union\cdots\union R_n \in \sh_1^\star$ and
 $T_1 \union\cdots\union T_n \in \sh_2^\star$.
Thus $S \in \bin(\sh_1^\star, \sh_2^\star)$.

On the other hand, 
$S \in \bin(\sh_1^\star, \sh_2^\star)$ means that $S = R \union T$ where,
for some $k,l \in \Nset$,
$R_1, \ldots, R_k \in \sh_1$, and
$T_1, \ldots, T_l \in \sh_2$,
we have
$R = R_1 \union \cdots \union R_k$ and
$T = T_1 \union \cdots \union T_l$.
Let $n$ be the maximum of $\{k, l\}$ and suppose that,
for each $i,j \in \Nset$ where
$k+1 \leq i \leq n$ and
$l+1 \leq j \leq n$, we define $R_i \defeq R_k$ and $T_j \defeq T_l$.
Then,
$S = (R_1 \union T_1) \union\cdots\union (R_n \union T_n)$.
However, for $1\leq i \leq n$, $R_i \union T_i \in \bin(\sh_1,\sh_2)$.
Thus $S \in  \bin(\sh_1, \sh_2)^\star$.
\end{proof}

\begin{proof*}[Proof of Theorem~\ref{thm: idempotence of amgu}.]
Let
\begin{align*}
  \sh_-
    &\defeq
      \irel(v_{xr}, \sh),\\
  \sh_{xr}
    &\defeq
      \bin\bigl(\rel(v_x, \sh)^\star, \rel(v_r, \sh)^\star\bigr).
\end{align*}
Then, by Lemma~\ref{lem:bin-star},
\begin{align*}
  \sh_{xr}^\star &= \sh_{xr},
    &
      \bin(\sh_{xr}, \sh_{xr}) &= \sh_{xr}. \\
\intertext{%
Moreover,
}
  \rel(v_x, \sh_{xr}) &= \sh_{xr},
    &
      \rel(v_x, \sh_-) &= \emptyset, \\
  \rel(v_r, \sh_{xr}) &= \sh_{xr},
    &
      \rel(v_r, \sh_-) &= \emptyset, \\
  \irel(v_{xr}, \sh_{xr}) &= \emptyset,
    &
      \irel(v_{xr}, \sh_-) &= \sh_-.
\end{align*}
Hence, we have
\begin{align*}
\rel(v_x, \sh_- \union \sh_{xr}) &= \sh_{xr},\\
\rel(v_r, \sh_- \union \sh_{xr}) &= \sh_{xr}, \\
\irel(v_{xr}, \sh_- \union \sh_{xr}) &= \sh_-.
\end{align*}

Now, by~(\ref{eq:amgu-in-terms-of-irel}),
\begin{align*}
  \amgu&\bigl(\amgu(\sh, x\mapsto r), x\mapsto r\bigr)\\
    &=
  \irel(v_{xr}, \sh_- \union \sh_{xr}) \union
 \bin\bigl(\rel(v_x, \sh_- \union \sh_{xr})^\star,
           \rel(v_r, \sh_- \union \sh_{xr})^\star
     \bigr)\\
    &=
  \sh_- \union \sh_{xr}\\
    &=
  \amgu(\sh, x\mapsto r).
\mathproofbox
\end{align*}
\end{proof*}

For the proof of commutativity, we require the following auxiliary
results.
\begin{lem}
\label{lem:irel-star-commute}
For each $V \in \wpf(\Vars)$ and $\sh \in \SH$ we have
\[
    \irel(V, \sh^\star) = \irel(V, \sh)^\star.
\]
\end{lem}
\begin{proof}
Let $S \in \SG$.
Then $S \in \irel(V, \sh^\star)$ means
$S \in \sh^\star$ and $S \inters V = \emptyset$.
In other words,
there exist $S_1$, \dots,~$S_n \in \sh$
such that $S = \bigunion_{i=1}^n S_i$ and, for each $i=1$, \dots,~$n$,
we have $S_i \inters V = \emptyset$.
This amounts to saying that there exist
$S_1$, \dots,~$S_n \in \irel(V, \sh)$ such that
$S = \bigunion_{i=1}^n S_i$,
which is equivalent to $S \in \irel(V, \sh)^\star$.
\end{proof}

The auxiliary function $\rel$ possesses a weaker property.
\begin{lem}
\label{lem:rel-star-semicommute}
For each $V \in \wpf(\Vars)$ and $\sh \in \SH$ we have
\[
    \rel(V, \sh^\star) \Sseq \rel(V, \sh)^\star.
\]
\end{lem}
\begin{proof}
Let $S \in \SG$.
Then $S \in \rel(V, \sh)^\star$ means that
there exist $S_1$, \dots,~$S_n \in \sh$
such that $S_i \inters V \neq \emptyset$, for each $i=1$, \dots,~$n$,
and $S = \bigunion_{i=1}^n S_i$.
Thus $S \inters V \neq \emptyset$ and $S \in \rel(V, \sh^\star)$.
Hence, $\rel(V, \sh^\star) \Sseq \rel(V, \sh)^\star$.
\end{proof}

\begin{lem}
\label{lem:rel-union-star}
For each $V \in \wpf(\Vars)$, $\sh_1, \sh_2 \in \SH$,
and $S \in \wpf(\Vars)$ we have
\begin{multline*}
  S \in \rel(V, \sh_1 \union \sh_2)^\star \union \{\emptyset\} \\
    \iff
      \exists S_1 \in \rel(V, \sh_1)^\star \union \{\emptyset\} \st
        \exists S_2 \in \rel(V, \sh_2)^\star \union \{\emptyset\} \st
          S = S_1 \union S_2.
\end{multline*}
\end{lem}
\begin{proof}
If $S = \emptyset$ the statement is trivial.

Suppose $S \in \rel(V, \sh_1\union \sh_2)^\star$.
Then,
for some $n \in \Nset$,
there exists $n$ sets $R_1, \ldots, R_n \in (\sh_1\union \sh_2)$
such that $R_i \inters V \neq \emptyset$ for each $i = 1$, \dots,~$n$,
and $S = \bigunion_{i=1}^n R_i$.
Suppose
\(
  S_j = \bigcup \{\,
                  R_i \in \sh_j
                \mid
                  1 \leq i \leq n
                \,\}
\)
for $j = 1$,~$2$.
Thus we have $S_1 \in \rel(V, \sh_1)^\star \union \{\emptyset\}$,
$S_2 \in \rel(V, \sh_2)^\star \union \{\emptyset\}$,
and $S = S_1 \union S_2$.

Suppose
\[
      \exists S_1 \in \rel(V, \sh_1)^\star \union \{\emptyset\} \st
        \exists S_2 \in \rel(V, \sh_2)^\star \union \{\emptyset\} \st
          S = S_1 \union S_2,
\]
with $S_1$ and $S_2$ not both empty.
Then,
for some $m \geq 0$ and $n \geq 0$,
there exist $R_1, \ldots, R_m \in \rel(V, \sh_1)$
and $T_1, \ldots, T_n \in \rel(V, \sh_2)$
such that $S_1 = \bigunion_{i=1}^m R_i$
and $S_2 = \bigunion_{i=1}^n T_i$.
Then $R_1, \ldots, R_m, T_1, \ldots, T_n \in \rel(V, \sh_1 \union \sh_2)$
and
\[
  S =
    \Bigl( \bigunion_{i=1}^m R_i \Bigr)
      \union
        \Bigl( \bigunion_{i=1}^n T_i \Bigr).
\]
Thus $S \in \rel(V, \sh_1 \union \sh_2)^\star$.
\end{proof}

\pagebreak[3]
\begin{lem}
\label{lem:rel-irel}
For each $V_1,V_2 \in \wpf(\Vars)$ and $\sh \in \SH$ we have
\[
  \rel\bigl(V_1,\irel(V_2,\sh)\bigr) = \irel\bigl(V_2, \rel(V_1,\sh)\bigr).
\]
\end{lem}
\begin{proof}
Suppose $S \in \SG$.
Then $S \in \rel\bigl(V_1, \irel(V_2, \sh)\bigr)$ means
$S \inters V_1 \neq \emptyset$ and $S \inters V_2 = \emptyset$.
Similarly,
$S \in \irel\bigl(V_2, \rel(V_1, \sh)\bigr)$ means that
$S \inters V_2 = \emptyset$ and $S \inters V_1 \neq \emptyset$.
\end{proof}

\begin{proof*}[Proof of Theorem~\ref{thm: commutativity of amgu}.]
We let $R$, $S$, $T$, and $U$ (possibly subscripted) denote
elements of $\sh^\star$.
The subscripts reflect certain properties of the sets.
In particular, subscripts $x,r,xr,y,t,yt$ indicate
sets of variables that definitely have
a variable in common with the subscripted set.
For example, $R_x$ is a set in $\sh^\star$ that
has a common element with $v_x$ and $T_{xt}$ is a set in $\sh^\star$ that
has common elements with $v_x$ and $v_t$.
In contrast,
the  subscript `$-$' indicates that the subscripted set
does not share with one of the sets $v_{xr}$ or $v_{yt}$.
Of course, in the proof, each set is
formally defined as needed.

Suppose that
\begin{align*}
  S &\in \amgu\bigl(\amgu(\sh, x \mapsto r), y\mapsto t\bigr).
\intertext{%
We will show that
}
  S &\in \amgu\bigl(\amgu(\sh, y \mapsto t), x \mapsto r\bigr).
\end{align*}
The converse then holds by simply exchanging $x$ and $y$, and
$r$ and $t$.

There are two cases due to the two components of the definition of $\amgu$
in Eq.~(\ref{eq:amgu-in-terms-of-irel}).

\paragraph*{Case 1.}
Assume
\[
  S \in \irel\bigl(v_{yt}, \amgu(\sh, x \mapsto r)\bigr).
\]
Then $S \in \amgu(\sh,x\mapsto r)$ and $S \inters v_{yt} = \emptyset$.
Again there are two possibilities.

\paragraph*{Subcase 1a.}
Suppose first that
\begin{align*}
    S &\in \irel(v_{xr}, \sh). \\
\intertext{%
Thus $S \in \sh$, and,
since in this case we have $S \inters v_{yt} = \emptyset$,
}
    S &\in \irel(v_{yt}, \sh). \\
\intertext{%
The alternative definition of $\amgu$, (\ref{eq:amgu-in-terms-of-irel}),
implies $\irel(v_{yt}, \sh) \sseq \amgu(\sh,y\mapsto t)$
and thus we have also
}
    S &\in \amgu(\sh, y \mapsto t). \\
\intertext{%
Now, since the hypothesis of this subcase
implies $S \inters v_{xr} = \emptyset$,
we obtain
}
    S &\in \irel\bigl(v_{xr}, \amgu(\sh, y \mapsto t)\bigr). \\
\intertext{%
Hence, again by~(\ref{eq:amgu-in-terms-of-irel}),
we can conclude that
}
    S &\in \amgu\bigl(\amgu(\sh, y \mapsto t),x \mapsto r\bigr).
\end{align*}

\paragraph*{Subcase 1b.}
Suppose now that
\[
  S\in \bin\bigl(\rel(v_x, \sh)^\star,\rel(v_r, \sh)^\star\bigr).
\]
Then, there exist $S_x, S_r \in \SG$
such that $S = S_x \union S_r$,
where
\begin{align*}
    S_x &\in \rel(v_x, \sh)^\star, &
    S_r &\in \rel(v_r, \sh)^\star. \\
\intertext{%
  By the hypothesis for this case we have $S \inters v_{yt} = \emptyset$
  and thus $S_x \inters v_{yt} = \emptyset$
  and $S_r \inters v_{yt} = \emptyset$.
  This allows to state that
}
    S_x &\in \irel\bigl(v_{yt}, \rel(v_x, \sh)^\star\bigr), &
    S_r &\in \irel\bigl(v_{yt}, \rel(v_r, \sh)^\star\bigr), \\
\intertext{%
and hence, by Lemma~\ref{lem:irel-star-commute},
}
    S_x &\in \irel\bigl(v_{yt}, \rel(v_x, \sh)\bigr)^\star, &
    S_r &\in \irel\bigl(v_{yt}, \rel(v_r, \sh)\bigr)^\star, \\
\intertext{%
Thus, by Lemma~\ref{lem:rel-irel},
}
  S_x &\in \rel\bigl(v_x, \irel(v_{yt}, \sh)\bigr)^\star, &
  S_r &\in \rel\bigl(v_r, \irel(v_{yt}, \sh)\bigr)^\star, \\
\intertext{%
so that,  by~(\ref{eq:emi-star}), (\ref{eq:emi-rel}),
and~(\ref{eq:amgu-in-terms-of-irel}),
}
    S_x &\in \rel\bigl(v_x, \amgu(\sh, y \mapsto t)\bigr)^\star, &
    S_r &\in \rel\bigl(v_r, \amgu(\sh, y \mapsto t)\bigr)^\star.
\end{align*}
Therefore,
\begin{align*}
  S_x \union S_r
    &\in
      \bin\Bigl(\rel\bigl(v_x, \amgu(\sh, y \mapsto t)\bigr)^\star,
                \rel\bigl(v_r, \amgu(\sh, y \mapsto t)\bigr)^\star\Bigr)\\
\intertext{%
so that, as $S_x \union S_r = S$, it follows from
(\ref{eq:amgu-in-terms-of-irel}) that
}
S
   & \in
     \amgu\bigl(\amgu(\sh, y \mapsto t), x \mapsto r\bigr).
\end{align*}

\paragraph*{Case 2.}
Assume
\[
  S \in \bin\Bigl(\rel\bigl(v_y, \amgu(\sh, x \mapsto r)\bigr)^\star,
                  \rel\bigl(v_t, \amgu(\sh, x \mapsto r)\bigr)^\star\Bigr).
\]
Then there exist $S_y, S_t \in \SG$ such that
\begin{align}
\label{eq:S-split}
    S &= S_y \union S_t \\
\intertext{%
  where
}
\label{eq:S-yt-def}
\begin{split}
    S_y &\in \rel\bigl(v_y, \amgu(\sh, x \mapsto r)\bigr)^\star, \\
    S_t &\in \rel\bigl(v_t, \amgu(\sh, x \mapsto r)\bigr)^\star.
\end{split} 
\end{align}
Then, by Lemma~\ref{lem:rel-star-semicommute},
\begin{align}
\label{eq:S-yt-v-yt-non-empty}
S_y \inters v_y \neq \emptyset,\qquad
 &S_t \inters v_t \neq \emptyset.\\
\intertext{%
By~(\ref{eq:amgu-in-terms-of-irel}) and Lemma~\ref{lem:rel-union-star},
  there exist $R_-$, $R_{xr}$, $T_-$, and $T_{xr}$ such that
}
\label{eq:R-xr-S}
    S_y = R_- \union R_{xr},\qquad
    &S_t = T_- \union T_{xr}
\end{align} 
  where
\begin{align}
\label{eq:RT-xryt-def}
\begin{split}
  R_-
    &\in
      \rel\bigl(v_y, \irel(v_{xr},\sh)\bigr)^\star \union \{\emptyset\}, \\
  R_{xr}
    &\in
      \rel\Bigl(v_y, \bin\bigl(\rel(v_x, \sh)^\star,
                               \rel(v_r, \sh)^\star\bigr)\Bigr)^\star
        \union \{\emptyset\}, \\
  T_-
    &\in
      \rel\bigl(v_t, \irel(v_{xr}, \sh)\bigr)^\star \union \{\emptyset\},\\
  T_{xr}
    &\in
      \rel\Bigl(v_t, \bin\bigl(\rel(v_x, \sh)^\star,
                               \rel(v_r, \sh)^\star\bigr)\Bigr)^\star
        \union \{\emptyset\}.
\end{split} \\
\intertext{%
  Then, by Lemmas~\ref{lem:rel-irel} and~\ref{lem:irel-star-commute},
}
\label{eq:RT-yt-noxr}
\begin{split}
  R_-
    &\in
      \irel\bigl(v_{xr}, \rel(v_y, \sh)^\star\bigr) \union \{\emptyset\}, \\
  T_-
    &\in
      \irel\bigl(v_{xr}, \rel(v_t, \sh)^\star\bigr) \union \{\emptyset\}.
\end{split} \\
\intertext{%
  Also, using Lemmas~\ref{lem:rel-star-semicommute}, \ref{lem:bin-star},
  and then the idempotence of $(\cdot)^\star$,
}
\label{eq:RT-xr-yt-in-bin}
\begin{split}
  R_{xr}
    &\in
      \rel\Bigl(v_y, \bin\bigl(\rel(v_x, \sh)^\star,
                               \rel(v_r, \sh)^\star\bigr)\Bigr)
        \union \{\emptyset\}, \\
  T_{xr}
    &\in
      \rel\Bigl(v_t, \bin\bigl(\rel(v_x, \sh)^\star,
                               \rel(v_r, \sh)^\star\bigr)\Bigr)
        \union \{\emptyset\}.
\end{split}
\end{align}

\paragraph*{Subcase 2a.}
Suppose $R_{xr} = T_{xr} = \emptyset$.
Then,
by~(\ref{eq:R-xr-S}),
\begin{align}
\label{eq:Sy-St=R-T-}
S_y = R_-,\quad
S_t = T_-.
\end{align}
By~(\ref{eq:S-yt-v-yt-non-empty}),
$R_-,T_- \neq \emptyset$
and hence, using~(\ref{eq:RT-yt-noxr}),
\begin{align*}
  R_- \union T_-
    &\in \bin\bigl(\rel(v_y, \sh)^\star, \rel(v_t, \sh)^\star\bigr), \\
\intertext{%
so that, by~(\ref{eq:amgu-in-terms-of-irel}),
}
  R_- \union T_- 
    &\in \amgu(\sh, y \mapsto t). \\
\intertext{%
Also, it follows from~(\ref{eq:RT-yt-noxr}) that
$R_- \inters v_{xr} = \emptyset$ and $T_- \inters v_{xr} = \emptyset$, so that
}
  R_- \union T_- 
    &\in \irel\bigl(v_{xr}, \amgu(\sh, y \mapsto t)\bigr).
\end{align*}
However, by~(\ref{eq:S-split}) and (\ref{eq:Sy-St=R-T-}),
$S= R_- \union T_-$ so that,
by~(\ref{eq:amgu-in-terms-of-irel}),
\[
 S \in \amgu\bigl(\amgu(\sh, y \mapsto t), x \mapsto r\bigr).
\]

\paragraph*{Subcase 2b.}
Suppose
$R_{xr} \union T_{xr} \neq \emptyset.$
Then, by~(\ref{eq:RT-xr-yt-in-bin}),
\begin{equation}
\label{eq:subcase-2b}
 (R_{xr} \union T_{xr}) \inters v_{yt} \neq \emptyset.
\end{equation}
The proof of this subcase is in two parts. In the
first part we divide $R_{xr}$ and $T_{xr}$ into a number of subsets.
In the second part, these subsets will be reassembled so as to prove
the required result.

First, by~(\ref{eq:RT-xr-yt-in-bin}), there
exist $R_x, R_r, T_x, T_r \in \wpf(\Vars)$ such that
\begin{align}
\label{eq:Rxr-Rx-Rr}
  R_{xr} &= R_x  \union R_r,
    & T_{xr} &= T_x  \union T_r,\\
\intertext{%
where either $R_x = R_r = \emptyset$ or
}
\notag
  R_x &\in \rel(v_x, \sh)^\star,
    & R_r &\in \rel(v_r, \sh)^\star,\\
\intertext{%
and either $T_x = T_r = \emptyset$ or
}
\notag
  T_x &\in \rel(v_x, \sh)^\star,
    & T_r &\in \rel(v_r, \sh)^\star.
\end{align}
Thus, if either $R_x \union T_x = \emptyset$ or  $R_r \union T_r = \emptyset$,
it follows that
\[
  R_{xr}  \union T_{xr}
    = (R_x \union R_r) \union (T_x \union T_r)
      = \emptyset.
\]
However, by~(\ref{eq:subcase-2b}),
$R_{xr}  \union T_{xr} \neq  \emptyset$, so that we have
\begin{align}
\label{eq:RT-xr-nonempty}
  R_x \union T_x &\neq \emptyset,
    & R_r \union T_r &\neq \emptyset.
\end{align}
We now subdivide the sets $R_x$, $T_x$, $R_r$, and $T_r$ further.
First note that
\begin{align*}
  \sh
    &=
      \irel(v_{yt}, \sh)
        \union \rel(v_y, \sh)
        \union \irel\bigl(v_y, \rel(v_t, \sh)\bigr), \\
  \sh
    &=
      \irel(v_{yt}, \sh)
        \union \irel\bigl(v_t, \rel(v_y, \sh)\bigr)
        \union\rel(v_t, \sh).
\end{align*}
Hence, by Lemma~\ref{lem:rel-union-star}, sets
$R_{x-}$, $R_{xy}$, $R_{xt}$,
$R_{r-}$, $R_{ry}$, $R_{rt}$,
$T_{x-}$, $T_{xy}$, $T_{xt}$,
$T_{r-}$, $T_{ry}$, $T_{rt} \in \wpf(\Vars)$ exist such that
\begin{equation}
\label{eq:RT-xr-split-on-yt}
\begin{aligned}
  R_x &= R_{x-} \union R_{xy} \union R_{xt}, \\
  R_r &= R_{r-} \union R_{ry} \union R_{rt},
\end{aligned}
\qquad
\begin{aligned}
  T_x &= T_{x-} \union T_{xy} \union T_{xt}, \\
  T_r &= T_{r-} \union T_{ry} \union T_{rt},
\end{aligned}
\end{equation}
where
\begin{align}
\label{eq:RT-xr-noyt-split}
\begin{split}
  R_{x-}, T_{x-} &\in \rel\bigl(v_x,\irel(v_{yt}, \sh)\bigr)^\star
    \union \{\emptyset\}, \\
  R_{r-}, T_{r-} &\in \rel\bigl(v_r,\irel(v_{yt}, \sh)\bigr)^\star
    \union \{\emptyset\},
\end{split} \\
\intertext{%
  and
}
\label{eq:RT-xr-yt-split}
\begin{split}
  R_{xy}, T_{xy} &\in \rel\bigl(v_x,\rel(v_y, \sh)\bigr)^\star
    \union \{\emptyset\},\\
  R_{ry}, T_{ry} &\in \rel\bigl(v_r,\rel(v_y, \sh)\bigr)^\star
    \union \{\emptyset\},\\
  R_{xt}, T_{xt} &\in \rel\bigl(v_x,\rel(v_t, \sh)\bigr)^\star
    \union \{\emptyset\}, \\
  R_{rt}, T_{rt} &\in \rel\bigl(v_r,\rel(v_t, \sh)\bigr)^\star
    \union \{\emptyset\},
\end{split}
\end{align}
and also
\begin{equation}
\label{eq:maxcond}
\begin{aligned}
  (R_x\setdiff R_{xy})  \inters v_y &= \emptyset, \\
  (R_r\setdiff R_{ry})  \inters v_y &= \emptyset,
\end{aligned}
\qquad
\begin{aligned}
  (T_x\setdiff T_{xt}) \inters v_t &= \emptyset, \\
  (T_r\setdiff T_{rt}) \inters v_t &= \emptyset.
\end{aligned}
\end{equation}
We note a few simple but useful consequences of these definitions.
First, it follows from~(\ref{eq:RT-xr-noyt-split})
using~(\ref{eq:emi-star}), (\ref{eq:emi-rel}),
and~(\ref{eq:amgu-in-terms-of-irel}),
that
\begin{equation}
\label{eq:RT-xr-noyt-rel-amgu}
\begin{split}
  R_{x-}, T_{x-}
    & \in \rel\bigl( v_x, \amgu(\sh, y \mapsto t)\bigr)^\star
      \union \{\emptyset\}, \\
  R_{r-}, T_{r-}
    & \in \rel\bigl( v_r, \amgu(\sh, y \mapsto t)\bigr)^\star
      \union \{\emptyset\}.
\end{split}
\end{equation}
Secondly, using~(\ref{eq:RT-xr-noyt-split}) with
Lemma~\ref{lem:rel-star-semicommute}, we have
\begin{equation}
\label{eq:RT-xr-noyt-in-irel-star}
  R_{x-}, T_{x-}, R_{r-}, T_{r-} \in
       \irel(v_{yt}, \sh)^\star \union \{\emptyset\},
\end{equation}
and then, using this with~(\ref{eq:subcase-2b}),
(\ref{eq:Rxr-Rx-Rr}), and (\ref{eq:RT-xr-split-on-yt}),
it follows that
\begin{equation}
\label{eq:RT-xr-yt-union-nonempty}
  R_{xy} \union T_{xy} \union R_{ry} \union T_{ry} \union
  R_{xt} \union T_{xt} \union R_{rt} \union T_{rt}
    \neq \emptyset.
\end{equation}

In the second part of the proof for this subcase, the component subsets
of $S$ are reassembled in an order that proves
the required result.
First, let
\begin{align}
\label{eq:U-yt-def}
\begin{split}
  U_y &\defeq R_- \union R_{xy} \union R_{ry} \union T_{xy} \union T_{ry}, \\
  U_t &\defeq T_- \union R_{xt} \union R_{rt} \union T_{xt} \union T_{rt}, 
\end{split}\\
\intertext{%
  and
}
\label{eq:U-split}
  U &\defeq U_y \union U_t. \\
\intertext{%
  By relations~(\ref{eq:RT-xryt-def})
  and~(\ref{eq:RT-xr-yt-split}) (with Lemma~\ref{lem:rel-star-semicommute}),
  each component set in the definition of $U_y$ is in
  $\rel(v_y, \sh)^\star \union \{\emptyset\}$ and each component set in the
  definition of $U_t$ is in $\rel(v_t,\sh)^\star \union \{\emptyset\}$. Thus,
  by the definition of $(\cdot)^\star$,
}
\notag
  U_y &\in \rel(v_y, \sh)^\star \union \{\emptyset\}, \\
  U_t &\in \rel(v_t, \sh)^\star \union \{\emptyset\}.
\end{align}
By~(\ref{eq:Rxr-Rx-Rr}) and (\ref{eq:maxcond}) we have
\begin{align*}
  \bigl(R_{xr} \setdiff (R_{xy} \union R_{ry})\bigr) \inters v_y
    &= \emptyset \\
\intertext{%
and hence, by~(\ref{eq:R-xr-S}), we have also that
}
  \bigl(S_y \setdiff (R_{xy} \union R_{ry} \union R_-)\bigr) \inters v_y
    &= \emptyset.
\end{align*}
By~(\ref{eq:S-yt-v-yt-non-empty}), $S_y \inters v_y \neq \emptyset$.
Thus, $R_{xy} \union R_{ry} \union R_- \neq \emptyset$ and,
as a consequence of~(\ref{eq:U-yt-def}), $U_y \neq \emptyset$.
For similar reasons, $U_t\neq \emptyset$. Hence, by~(\ref{eq:U-split}),
\begin{align}
\notag
  U
    &\in
      \bin\bigl(\rel(v_y, \sh)^\star,
                \rel(v_t, \sh)^\star\bigr), \\
\intertext{%
and therefore, using~(\ref{eq:amgu-in-terms-of-irel}), it follows that
}
\label{eq:U-in-amgu}
  U &\in \amgu(\sh,y \mapsto t).
\end{align}
Now, by~(\ref{eq:RT-xr-yt-union-nonempty}),
at least one of the following two inequalities holds:
\begin{align}
\label{eq:inequalities}
\begin{split}
  R_{xy} \union T_{xy} \union R_{xt} \union T_{xt} &\neq \emptyset,\\
  R_{ry} \union T_{ry} \union R_{rt} \union T_{rt} &\neq \emptyset.
\end{split}
\intertext{%
Assume first that
  $R_{xy} \union T_{xy} \union R_{xt} \union T_{xt} = \emptyset$ and
  $R_{ry} \union T_{ry} \union R_{rt} \union T_{rt} \neq \emptyset.$
Then, using~(\ref{eq:RT-xr-nonempty}) and~(\ref{eq:RT-xr-split-on-yt})
with the first of these,
}
\notag
  R_{x-} \union T_{x-} &\neq \emptyset.\\
\intertext{%
Also, using~(\ref{eq:RT-xr-yt-split}) with the second, we have
   $(R_{ry} \union R_{rt} \union T_{ry} \union T_{rt})
   \inters v_r \neq \emptyset$
and therefore it follows from~(\ref{eq:U-yt-def}) and (\ref{eq:U-split}),
that
}
\notag
  U \inters v_r &\neq \emptyset.
\end{align}
Hence,
by~(\ref{eq:RT-xr-noyt-rel-amgu}) and~(\ref{eq:U-in-amgu}),
\begin{align}
\label{eq:1st-of-inequality-comb}
\begin{split}
  R_{x-} \union T_{x-}
    &\in
      \rel\bigl(v_x, \amgu(\sh, y \mapsto t)\bigr)^\star, \\
  U \union R_{r-} \union T_{r-}
    &\in
      \rel\bigl(v_r, \amgu(\sh, y \mapsto t)\bigr)^\star.
\end{split}
\end{align}
Similarly, assuming
  $R_{xy} \union T_{xy} \union R_{xt} \union T_{xt} \neq \emptyset$ and
  $R_{ry} \union T_{ry} \union R_{rt} \union T_{rt} = \emptyset$
 it follows that
\begin{align}
\label{eq:2nd-of-inequality-comb}
\begin{split}
  R_{r-} \union T_{r-}
    &\in
      \rel\bigl(v_r, \amgu(\sh, y \mapsto t)\bigr)^\star, \\
  R_{x-} \union T_{x-} \union U
    &\in
      \rel\bigl(v_x,\amgu(\sh, y \mapsto t)\bigr)^\star. 
\end{split}\\
\intertext{%
  Finally, assuming
  $R_{xy} \union T_{xy} \union R_{xt} \union T_{xt} \neq \emptyset$ and
  $R_{ry} \union T_{ry} \union R_{rt} \union T_{rt} \neq \emptyset$
  it follows from~(\ref{eq:RT-xr-yt-split}) that
  $U \inters v_x \neq \emptyset$ and  $U \inters v_r \neq \emptyset$,
  and hence
}
\label{eq:3rd-of-inequality-comb}
\begin{split}
  R_{x-} \union T_{x-} \union U
    &\in
      \rel\bigl(v_x, \amgu(\sh, y \mapsto t)\bigr)^\star, \\
  U \union R_{r-} \union T_{r-}
    &\in
      \rel\bigl(v_r, \amgu(\sh, y \mapsto t)\bigr)^\star.
\end{split}
\end{align}
Thus, as one of the inequalities in (\ref{eq:inequalities})
holds, one of (\ref{eq:1st-of-inequality-comb}),
(\ref{eq:2nd-of-inequality-comb}) 
or~(\ref{eq:3rd-of-inequality-comb}) holds so that
\begin{multline*}
  R_{x-} \union T_{x-} \union U \union R_{r-} \union T_{r-}\\
  \in
    \bin\Bigl(\rel\bigl(v_x, \amgu(\sh, y \mapsto t)\bigr)^\star,
              \rel\bigl(v_r, \amgu(\sh, y \mapsto t)\bigr)^\star\Bigr).
\end{multline*}
 However, since
\begin{align*}
  S &= R_{x-} \union T_{x-} \union U \union R_{r-} \union T_{r-}, \\
\intertext{%
we have
}
  S &\in
    \bin\Bigl(\rel\bigl(v_x, \amgu(\sh, y \mapsto t)\bigr)^\star,
              \rel\bigl(v_r, \amgu(\sh, y \mapsto t)\bigr)^\star\Bigr). \\
\intertext{%
  Hence, by~(\ref{eq:amgu-in-terms-of-irel}),
}
  S &\in \amgu\bigl(\amgu(\sh, y \mapsto t), x \mapsto r\bigr).
\mathproofbox
\end{align*}
\end{proof*}

\subsection{Proofs of Results for Sharing Domains}
\label{sec:aunify results}

We prove all the results in this section by induction on the
cardinality of a substitution $\nu$.
For each result, the proof is obvious if $\nu$ is empty or does not unify.
Thus, in the following proofs, we assume that $\nu$ unifies and is non-empty.
We suppose that $(x \mapsto r) \in \nu$
and let $\nu' \defeq \nu \setdiff \{x \mapsto r\}$.

\pagebreak[4]
\begin{proof*}[Proof of Lemma~\ref{lem: commutativity of aunify/Amgu}.]
We have
\begin{align*}
 \aunify&\Bigl(\Amgu\bigl((\sh,U), y\mapsto t\bigr), \nu\Bigr)
 \\
 &= \aunify\biggl(
         \Amgu\Bigl(\Amgu\bigl((\sh,U),y\mapsto t\bigr), x\mapsto r
              \Bigr), \nu'
        \biggr)
 &&\just{Def.~\ref{def:aunify}}\\
 &= \aunify\biggl(
         \Amgu\Bigl(\Amgu\bigl((\sh,U),x\mapsto r\bigr), y\mapsto t
              \Bigr), \nu'
        \biggr)
 &&\just{Cor.~\ref{cor: commutativity of Amgu}} \\
 &= \Amgu\biggl(
         \aunify\Bigl(\Amgu\bigl((\sh,U),x\mapsto r\bigr), \nu'
              \Bigr), y\mapsto t
        \biggr)
 &&\just{induction} \\
 &= \Amgu\Bigl(\aunify\bigl((\sh,U), \nu\bigr), y\mapsto t\Bigr)
 &&\just{Def.~\ref{def:aunify}}.
\mathproofbox
\end{align*}
\end{proof*}

\begin{proof*}[Proof of Theorem~\ref{thm: soundness of aunify}.]
Let $\mu'$ be a most general solution for $(\nu' \union \sigma)$. Then
\begin{align*}
  \alpha&(\sigma, U) \leqSSl (\sh,U)\\
  &\implies \alpha\bigl(\mu', U\union \vars(\nu')\bigr)
\\&\qquad\qquad\qquad\qquad
         \leqSSl
   \aunify\bigl((\sh,U), \nu'\bigr)
 &&\just{induction}\\
  &\implies \alpha\bigl(
          \mu, U \union \vars(\nu)
                  \bigr)
\\&\qquad\qquad\qquad\qquad
         \leqSSl
   \Amgu\Bigl(\aunify\bigl((\sh,U), \nu'\bigr), x\mapsto r\Bigr)
 &&\just{Cor.~\ref{cor: soundness of Amgu}} \\
  &\implies \alpha\bigl(
          \mu, U \union \vars(\nu)
                  \bigr)
\\&\qquad\qquad\qquad\qquad
         \leqSSl
   \aunify\Bigl(\Amgu\bigl((\sh,U),  x\mapsto r\bigr), \nu'\Bigr)
 &&\just{Lem.~\ref{lem: commutativity of aunify/Amgu}}\\
  &\implies \alpha\bigl(\mu, U \union \vars(\nu)\bigr)
\\&\qquad\qquad\qquad\qquad
         \leqSSl
   \aunify\bigl((\sh, U), \nu\bigr)
 &&\just{Def.~\ref{def:aunify}}.
\mathproofbox
\end{align*}
\end{proof*}

\begin{proof*}[Proof of Theorem~\ref{thm: idempotence of aunify}.]
We have
\begin{align*}
&\aunify\Bigl(\aunify\bigl((\sh,U), \nu\bigr), \nu\Bigr)\\
  &= \aunify\biggl(
         \Amgu\Bigl(\aunify\bigl(\Amgu((\sh,U),x\mapsto r), \nu'
                           \bigr), x\mapsto r
              \Bigr), \nu'
        \biggr)
 &&\just{Def.~\ref{def:aunify}}\\
  &= \aunify\biggl(
         \aunify\Bigl(\Amgu\bigl(\Amgu((\sh,U),x\mapsto r), x\mapsto r
                           \bigr), \nu'
              \Bigr), \nu'
         \biggr)
 &&\just{Lem.~\ref{lem: commutativity of aunify/Amgu}} \\
  &= \aunify\Bigl(\Amgu\bigl(\Amgu((\sh,U),x\mapsto r), x\mapsto r
                           \bigr), \nu'
              \Bigr)
 &&\just{induction} \\
  &= \aunify\Bigl(\Amgu\bigl((\sh,U),x\mapsto r
                     \bigr), \nu'
        \Bigr)
 &&\just{Cor.~\ref{cor: idempotence of Amgu}} \\
  &= \aunify\bigl((\sh,U), \nu\bigr)
 &&\just{Def.~\ref{def:aunify}}.
\mathproofbox
\end{align*}
\end{proof*}

\begin{proof*}[Proof of Theorem~\ref{thm: commutativity of aunify}.]
The induction is on the set of equations $\nu_1$.
The comments at the start of this section apply therefore to $\nu_1$ instead
of $\nu$ and thus we let
$\nu_1' \defeq \nu_1 \setdiff \{x \mapsto r\}$ so that we have
\begin{align*}
 \aunify&\Bigl(\aunify\bigl((\sh,U), \nu_1\bigr), \nu_2\Bigr)\\
 &= \aunify\biggl(
         \aunify\Bigl(\Amgu\bigl((\sh,U), x\mapsto r\bigr), \nu_1'
              \Bigr), \nu_2
        \biggr)
 &&\just{Def.~\ref{def:aunify}}\\
 &= \aunify\biggl(
         \aunify\Bigl(\Amgu\bigl((\sh,U), x\mapsto r\bigr), \nu_2
              \Bigr), \nu_1'
        \biggr)
 &&\just{induction}\\
 &= \aunify\biggl(
         \Amgu\Bigl(\aunify\bigl((\sh,U), \nu_2\bigr), x\mapsto r
              \Bigr), \nu_1'
        \biggr)
 &&\just{Lem.~\ref{lem: commutativity of aunify/Amgu}} \\
 &= \aunify\Bigl(\aunify\bigl((\sh,U), \nu_2\bigr), \nu_1\Bigr)
 &&\just{Def.~\ref{def:aunify}}.
\mathproofbox
\end{align*}
\end{proof*}

\section{Conclusion}
\label{sec: discuss}

The $\Sharing$ domain,
which was defined in~\cite{JacobsL89,Langen90th},
is considered to be the principal abstract domain
for sharing analysis of logic programs
in both practical work and theoretical study.
For many years, this domain was accepted and implemented as it was.
However, in \cite{BagnaraHZ97b}, we proved that
$\Sharing$ is, in fact, redundant for pair-sharing and
we identified the weakest abstraction of $\Sharing$
that can capture pair-sharing with the same degree of precision.
One notable advantage of this abstraction is that
the costly star-union operator is no longer necessary.
The question of whether the abstract operations for $\Sharing$ were
complete or optimal was studied by Cortesi and Fil\'e~\cite{CortesiF99}.
Here it is proved that although the `$\sqcup$' and projection operations
are complete (and hence, optimal), $\aunify$ is optimal but not complete.
The problem of scalability of $\Sharing$,
still retaining as much precision as possible,
was tackled in \cite{ZaffanellaBH99},
where a family of widenings is presented
that allow the desired goal to be achieved.
In \cite{ZaffanellaHB99,ZaffanellaHB01TPLP},
the decomposition of $\Sharing$ and
its non-redundant counterpart via complementation is studied.
This shows the close relationship between these domains
and $\PS$ (the usual domain for pair-sharing)
and $\Def$ (the domain of definite Boolean functions).
Many sharing analysis techniques and/or enhancements
have been advocated to have potential
for improving the precision of the sharing information
over and above that obtainable using the classical combination
of $\Sharing$ with the usual domains for linearity and freeness.
Moreover, these enhancements had been circulating for years
without an adequate supporting experimental evaluation.
Thus we investigated these techniques to see if and by how much
they could improve precision.
Using the \china{} analyzer~\cite{Bagnara97th}
for the experimental part of the work, we discovered that,
apart from the enhancement that upgrades $\Sharing$
with structural information, these techniques had little
impact on precision~\cite{BagnaraZH00}.

In this paper, we have defined a new abstraction function
mapping a set of substitutions in rational solved form
into their corresponding sharing abstraction.
The new function is a generalisation of the classical
abstraction function of \cite{JacobsL89},
which was defined for idempotent substitutions only.
Using our new abstraction function,
we have proved the soundness of the classical
abstract unification operator $\aunify$.
Other contributions of our work are the formal proofs of
the commutativity and idempotence of the $\aunify$ operator
on the $\Sharing$ domain.
Even if commutativity was a known property,
the corresponding proof in \cite{Langen90th} was not satisfactory.
As far as idempotence is concerned, our result differs from
that given in \cite{Langen90th}, which was based on a composite
abstract unification operator performing also the renaming
of variables.
It is our opinion that our main result, the soundness of
the $\aunify$ operator, is really valuable
as it allows for the safe application
of sharing analysis based on $\Sharing$
to any constraint logic language supporting
syntactic term structures,
based on either finite trees or rational trees.
This happens because our result does not rely
on the presence (or even the absence) of the occurs-check
in the concrete unification procedure implemented
by the analysed language.
Furthermore, as the groundness domain $\Def$ is included in $\Sharing$, 
our main soundness result also shows that $\Def$ is sound for 
non-idempotent substitutions.

From a technical point of view, we have introduced
a new class of concrete substitutions based on the notion 
of \emph{variable-idempotence}, generalizing the classical
concept of idempotence.
We have shown that any substitution is equivalent
to a variable-idempotent one, providing a finite sequence of transformations
for its construction.
This result assumes an arbitrary equality theory
and is therefore applicable to the study of any
abstract property which is preserved by logical equivalence.
Our application of this idea to the study of the soundness
of abstract unification for $\Sharing$ has shown that
it is particularly suitable for data-flow analyzers
where the corresponding abstraction function only depends on
the set of variables occurring in a term.
However, we believe that this concept can be usefully
exploited in a more general context.
Possible applications include
the proofs of optimality and completeness of abstract operators
with respect to the corresponding concrete operators
defined on a domain of substitutions in rational solved form.


\hyphenation{ Bie-li-ko-va Bruy-noo-ghe Common-Loops DeMich-iel Dober-kat
  Er-vier Fa-la-schi Fell-eisen Gam-ma Gem-Stone Glan-ville Gold-in Goos-sens
  Graph-Trace Grim-shaw Her-men-e-gil-do Hoeks-ma Hor-o-witz Kam-i-ko Kenn-e-dy
  Kess-ler Lisp-edit Lu-ba-chev-sky Nich-o-las Obern-dorf Ohsen-doth Par-log
  Para-sight Pega-Sys Pren-tice Pu-ru-sho-tha-man Ra-guid-eau Rich-ard Roe-ver
  Ros-en-krantz Ru-dolph SIG-OA SIG-PLAN SIG-SOFT SMALL-TALK Schee-vel
  Schlotz-hauer Schwartz-bach Sieg-fried Small-talk Spring-er Stroh-meier
  Thing-Lab Zhong-xiu }

\end{document}